\documentclass[12pt,preprint]{aastex}
\usepackage{emulateapj5}

\def\nop{\noindent}

\begin{document}

\title{Constraints in Cosmological Parameter Space from  \\
          the Sunyaev-Zel'dovich Effect and \\
           Thermal Bremsstrahlung}

\author{S. M. Molnar\altaffilmark{1,2}, M. Birkinshaw\altaffilmark{3}, 
        and R. F. Mushotzky\altaffilmark{1}}

\altaffiltext{1}{NASA/Goddard Space Flight Center, 
Laboratory for High Energy Astrophysics, Greenbelt, MD 20771}
\altaffiltext{2}{Present address: Department of Physics and Astronomy, 
Rutgers University, 136 Frelinghuysen Road, Piscataway, NJ 08854, 
sandorm@physics.rutgers.edu}
\altaffiltext{3}{Department of Physics, Bristol University, Tyndall Avenue, 
Bristol, BS8 1TL, UK}

\begin{abstract}

\nop  
We discuss how the space of possible cosmological parameters is
constrained by the angular diameter distance function, $D_A(z)$, as
measured using the SZ/X-ray method which combines Sunyaev-Zel'dovich
(SZ) effect and X-ray brightness data for clusters of galaxies.
New X-ray satellites, and ground-based interferometers dedicated to SZ
observations, should soon lead to $D_A(z)$ measurements limited by 
systematic rather than random error. We analyze the systematic and
random error budgets to make a realistic estimate of the accuracy
achievable in the determination of $(\Omega_m,\Omega_\Lambda,h)$, 
the density parameters of matter and cosmological constant, and the 
dimensionless Hubble constant, using $D_A(z)$ derived from the SZ/X-ray
method, and the position of the first ``Doppler'' peak in the cosmic 
microwave background fluctuations. 
We briefly study the effect of systematic errors. 
We find that $\Omega_m$, $\Omega_\Lambda$, and $w$ are affected, 
but $h$ is not by systematic errors which grow with redshift.
With as few as 70 clusters, each providing a measurement of $D_A(z)$
with a $7\%$ random and $5\%$ systematic error, 
$\Omega_m$ can be constrained to $\pm 0.2$, $\Omega_\Lambda$ to 
$\pm 0.2$, and $h$ to $\pm 0.11$ (all at $3\sigma$). 
We also estimate constraints for the alternative three-parameter set 
$(\Omega_m,w,h)$, where $w$ is the equation of state parameter. 
The measurement of $D_A(z)$ provides constraints complementary 
to those from the number density of clusters in redshift space. 
A sample of 70 clusters ($D_A$ measured with the same accuracy 
as before) combined with cluster evolution results (or a known matter 
density), can constrain $w$ within $\pm 0.45$ (at 3$\sigma$). 
Studies of X-ray and SZ properties of clusters of galaxies 
promise an independent and powerful test for cosmological parameters. 

\end{abstract}

% % % % % % % % % % % % % % % % % % % % % % % % % % % % % % % % % % % % % % % %
\section{Introduction}
\label{S:INTRO}

What set of cosmological parameters characterizes our Universe?

According to the most popular cold dark matter (CDM) scenario, the
Universe consists of baryonic matter and a substantial amount of
``dark'' matter. A variety of recent measurements have led to the
conclusion that the matter density parameter $\Omega_m \approx 0.3$
\citep{Turn00}, while CMB measurements strongly favor a flat space
time \citep{Bernet01,Lee_et01}, and SNe Ia measurements indicate that
the Universe is accelerating, suggesting a negative pressure
\citep{Rieset00,Perlet99}. Taken together, these pieces of evidence
suggest that the baryonic and dark matter content of the Universe is
supplemented by an additional smooth component with negative
pressure, $P_w$, modeled by the equation of state $\rho_w \, c^2 = -w
\, P_w$, where $\rho_w$ is the density of this component, $w$ is a
dimensionless state parameter of order unity (cf. Huterer \& Turner
2000).

Each existing dataset constrains, with limited accuracy, some subset
of the cosmological parameters. Different measurements and
combinations of measurements, such as SNe Ia, Cosmic Microwave Background 
(CMB) fluctuations, IRAS infrared galaxy surveys, classical double
radio galaxy properties, 1.2-Jy galaxy redshift surveys, gravitational
lensing, cluster X-ray temperature function and cluster number counts,
baryon and gas mass fraction, and the SZ effect have been used to
constrain cosmological parameters \citep{Jaffet00,Balbet00,TegmZald00,
Gueret00,Efstet99,Laseet99,Perlet99,GewiSilk98,Line98,Whit98,Pen_97,Sasa96,
HuteTurn00,MajuSubr00,Bridet99,Dieget01}.

In the future, the SNAP project (http://snap.lbl.gov) plans
to use the SNe Ia method to determine the matter density and the
cosmological constant at the few percent level. Even with the
next-generation CMB satellites, MAP and Planck, degeneracies will
remain among the cosmological parameters that can be estimated from
the results (Efstathiou \& Bond 1999; Zaldarriaga, Spergel \& Seljak 1997). 
The importance of using a wide range of
methods, therefore, is twofold. First, a simultaneous consideration of
all data sets should allow the best joint estimation of the
cosmological parameters. Second, the agreement of different techniques
for measuring the cosmological parameters should provide a cross-check
of our understanding of the underlying processes and a control against
systematic errors. As we extend our analysis of the CMB to more
complicated models (tensor fluctuations, finite neutrino masses, etc.)
the number of cosmological parameters increases, and it becomes even
more important that the widest possible range of datasets is used, and
that strong controls against systematic errors are in place.

Many of the techniques of cosmological parameter estimation use
clusters as tracer particles. As a result there is a large number of
planned cluster surveys in the two most important non-optical
observational indicators of clustering: the SZ effect and cluster
X-ray emission. Sunyaev-Zel'dovich effect surveys with dedicated
interferometers or receiver arrays will observe hundreds of clusters
with $z > 0.5$ per year (Browne et al. 2000; Holder, Carlstrom \&
Mohr 2000; Bartlett 2000). The new X-ray missions (Chandra, XMM) will 
provide data on hundreds of clusters with high redshift through their
deep and medium-deep surveys.

Cluster evolution, the redshift distribution of clusters from SZ 
and X-ray surveys, $N_{SZ}(z)$ and $N_X(z)$, and cluster number counts 
as a function of X-ray flux, $N_X(S)$, are important constraints on 
cosmological parameters. 
While methods based on the CMB power spectrum and SNe Ia are sensitive
to the angular diameter distance, cluster evolution (and number
counts) is sensitive to the growth function of matter density
fluctuations.
\cite{Bart00} estimated the performance of 
ground-based, arcminute-resolution, SZ surveys and concluded that more 
clusters will be detected with  deep, small-area surveys than shallow, 
wide-area surveys. \cite{Kneiet01} studied the performance of the
Arcminute MicroKelvin Imager experiment and showed that a set of only
about 20~clusters, with redshifts in the range 
$z =$ 0 - 0.8 is needed to measure $N_{SZ}(z)$ sufficiently well
to distinguish between $\Omega_m$ = 1 and 
$\Omega_m$ = 0.3 cosmologies. 
\cite{Carlet01} discuss a deep SZ ground based survey, and quantify 
constraints from $N_{SZ}(z)$ on $\Omega_m$ and $\Omega_\Lambda$. 
$N_{SZ}(S)$ and $N_{SZ}(z)$ were estimated from the proposed shallower, 
but all-sky Planck survey by \cite{Dieget01}, who concluded that about 
300 clusters (with the necessary optical follow-up to measure redshifts) 
would suffice to distinguish between open $\Omega_m$ = 0.3 and flat 
$\Omega_m$ = 1 cosmologies at $3\sigma$ confidence. 
Holder, Haiman \& Mohr (2001) discussed the constraints on the
parameter space defined by $(\Omega_m, \Omega_\Lambda, \sigma_8)$ 
(where $\sigma_8$ is the normalization of the matter power spectrum) 
using cluster evolution. 
Holder et al. showed that constraints from cluster evolution
and SNe Ia observations are highly complementary to each other. 
\cite{Haimet00} discussed the constraints on the
$(\Omega_m, w, h)$ parameter space, assuming a spatially flat geometry
($\Omega_\Lambda = 1 - \Omega_m$), that follow from an SZ effect
survey and a large angle deep X-ray survey (the Cosmology Explorer;
Ricker and Lamb). They found that $N_{SZ}(z)$ and $N_X(z)$, 
combined with constraints from CMB or SNe Ia experiments,
significantly reduce the degeneracies between $\Omega_m$, $w$, and $h$. 
\cite{HuteTurn00} estimated the constraints on $\Omega_m$
and $w$ for flat geometry that can be gained by combining results from
SNAP, Planck and SZ and X-ray surveys.

As has been realized, the angular diameter distance-redshift relation,
$D_A(z)$, is at the heart of many of these techniques, and is
sensitive to some important combinations of cosmological parameters
while being degenerate under others  
\citep{Jaffet00,TegmZald00,Efstet99,Laseet99,Perlet99,Whit98}.
Recently \cite{Whit98} estimated constraints on the pairs of quantities
$(\Omega_m,\Omega_\Lambda)$ and $(\Omega_m,w)$ (the latter in a flat
Universe) from the $D_A(z)$ function based on current SNe Ia data
combined with CMB first peak constraints. The analysis shows that the
constraint on parameters based on $D_A(z)$ is nearly orthogonal to
the constraint based on the position of the first peak in the
CMB fluctuation spectrum. These two datasets are thus highly
complementary, and form a particularly powerful pair of measurements
(see also Tegmark et al. 1998).

\begin{figure}
\figurenum{1}
\centerline{
\plotone{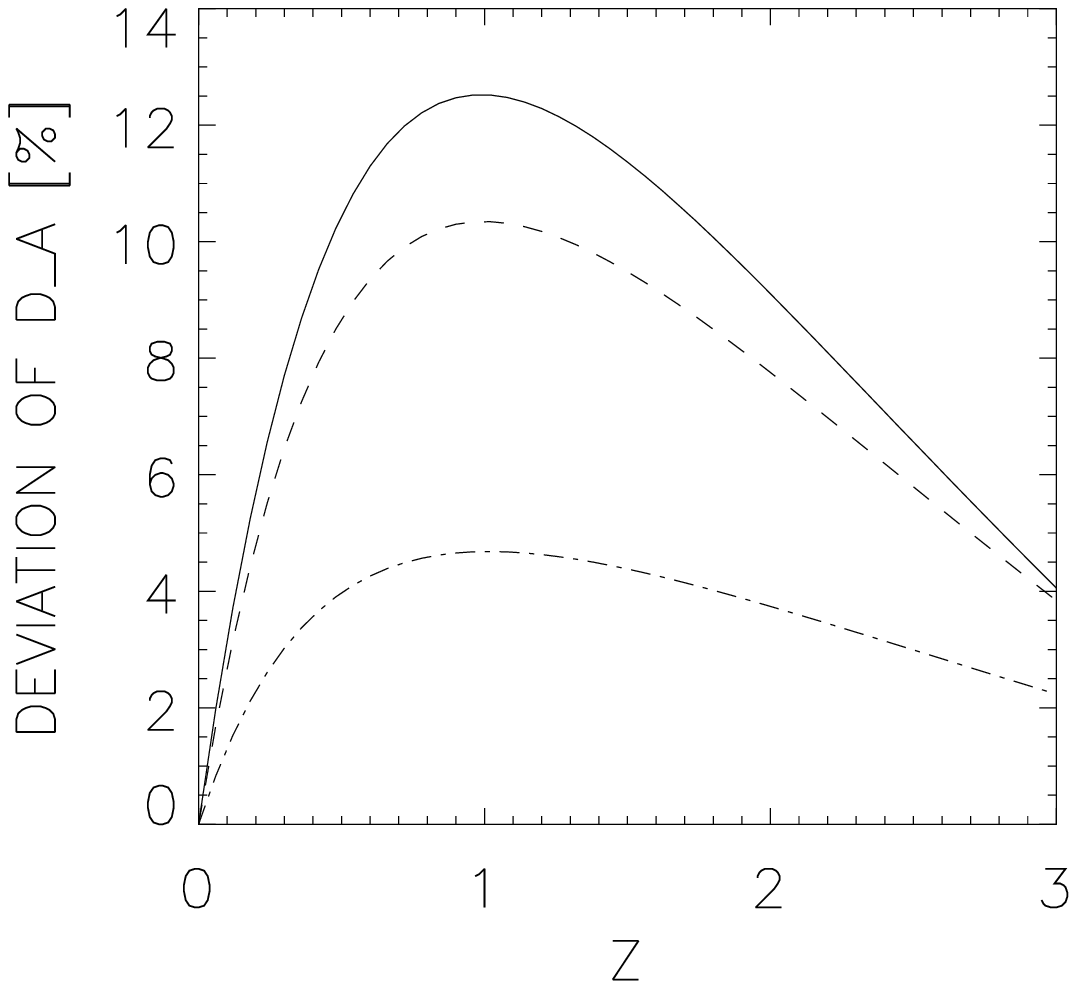}
}
\caption{
Deviation of the angular diameter distance in Universes with $\Omega_m
= 0.3$ and $\Omega_\Lambda =$ 0.7, 0.6, and 0.3 relative to the
angular distance function for a Universe with $\Omega_m = 0.3$ and
$\Omega_\Lambda = 0$ model, as a function of $z$ (with $h = 0.65$). 
The solid, short dashed, and dash dotted lines represent percentage deviations
with $\Omega_\Lambda =$ 0.7, 0.6, and 0.3 respectively.
\label{F:FIG1}
}
\end{figure}

The shape and normalization of the observed angular diameter distance
function constrains several cosmological parameters (the standard
formulae for distance in Friedmann-Robertson-Walker Universes are
given in, for example, Peebles 1993). The distance - redshift
function, $D_A(z)$, in CDM models depends on the matter density,
cosmological constant and Hubble constant, and any other particle
density which contributes to the curvature of space-time. The
slope of the distance-redshift function at low redshift is a measure
of the Hubble constant, while the shape of the function depends on the
curvature and the different densities. In Figure~\ref{F:FIG1} we show
the fractional difference in $D_A(z)$ with fixed matter density and
Hubble constant ($\Omega_m = 0.3$, $h = 0.65$), but various values
cosmological constants ($\Omega_\Lambda$ = 0.7, 0.6, 0.3, solid,
dashed and dash dotted lines) relative to a model with zero
cosmological constant ($\Omega_\Lambda = 0$). It can be seen from this
figure, that $D_A(z)$ is most sensitive to the value of the
cosmological constant at redshift about unity, and quite insensitive
to that at small or high redshifts. In the redshift interval from
$z=0.5$ to $z=1.8$ the angular diameter distance for flat
($\Omega_m=0.3$, $\Omega_\Lambda=0.7$) model is more than 10$\%$
different from a model with the same matter density but zero
cosmological constant ($\Omega_m=0.3$, $\Omega_\Lambda=0$).

We can expect high precision data from hundreds of clusters of
galaxies in the near future. With the present instrument suite, the
statistical errors on individual measurements will be small, and so
the usefulness of the data will be limited by their systematic errors.
In this paper we evaluate the error budget of distance determination
based on the SZ effect and X-ray measurements (assuming that the
X-ray output is dominated by thermal bremsstrahlung, as is appropriate
for the hot clusters in which the SZ effect is strong), and provide a
realistic estimate of errors achievable in the angular diameter
distance. We estimate how well one will be able to constrain the two
parameter sets $(\Omega_m, \Omega_\Lambda, h)$, and 
$(\Omega_m, w, h)$ (assuming a spatially flat geometry, $\Omega_\Lambda
= 1-\Omega_m$) if this $D_A(z)$ function is combined with the
position of the first Doppler peak in the angular power spectrum of
CMB fluctuations.

Our treatment is complementary to previous work of \cite{Kneiet01}, 
\cite{Carlet01}, \cite{Dieget01}, \cite{Holdet01}, \cite{Haimet00}, 
and \cite{HuteTurn00}, who used $N_{SZ}(z)$,
$N_{SZ}(S)$, and $N_{X}(z)$ to constrain cosmological parameters,
since we discuss the importance of $D_A(z)$. It also complements
the work by \cite{Whit98}, who used $D_A(z)$ determined from existing
SNe Ia data to constrain $(\Omega_m, \Omega_\Lambda)$, and
$(\Omega_m, w)$ (flat): the errors from the SZ/X-ray technique have
significantly different characteristics, and we are concerned with the
limitations that will be encountered with future survey data.

In the next section we briefly describe the well-known method of
angular distance determination based on the SZ effect and thermal
bremsstrahlung with an emphasis on how measurements are used, and how
their error propagate to the angular diameter distance. In
section~\ref{s:errors} we give a detailed analysis of the error
budget, in section~\ref{s:Constraints} we discuss the constraints on
cosmological parameters. Finally, section~\ref{S:Conclusion}
summarizes our conclusions.

% % % % % % % % % % % % % % % % % % % % % % % % % % % % % % % % % % % % % % % %
\section{Determination of Angular Diameter Distance using Clusters}

Distance determinations using the SZ effect and X-ray emission from
the intra-cluster medium (SZ/X method hereafter) are based on the fact
that these processes depend on different combinations of physical
parameters of the clusters. The SZ effect (Sunyaev and Zel'dovich
1980; for recent reviews cf. Birkinshaw 1999, and Rephaeli 1995) is a
result of inverse Compton scattering of CMB photons off hot electrons
in the intra-cluster (IC) gas. The number of photons is conserved, but,
on average, the photons gain energy, and thus generate a decrement in the
Rayleigh-Jeans part of the spectrum and an increment in the Wien
region. The amplitude of the SZ effect does not depend on the redshift
of the cluster.
We will discuss static thermal and kinematic thermal SZ effects in
this paper, where the ``static'' effect is present in all clusters,
and the ``kinematic'' effect is only present for those clusters with a
non-zero line-of-sight (LOS) peculiar velocity relative to the Hubble flow. 
A typical Rayleigh-Jeans decrement of the static SZ effect
is about about 50 times that of the kinematic SZ effect.
The static SZ effect is proportional to the LOS pressure integral of
the IC gas
\begin{equation} \label{E:SZ_1}
  \Delta T \propto  g(\nu) \int dz  \; n_e({\bf r}) \; T_e({\bf r})
  ,
\end{equation}
where $g(\nu)$ is the frequency dependence of the effect ($g
\rightarrow -2$ in the non-relativistic Rayleigh-Jeans limit), and
$n_e({\bf r})$ and $T_e({\bf r})$ are the electron density and
temperature as functions of position within the cluster. The central
X-ray surface brightness of a cluster is emission-weighted
line-of-sight average of $n_e(z)^2 T_e(z)$:
\begin{equation} \label{E:THBR_1}
  S_X \propto { 1 \over (1+z)^4} \int dz\; n_e(z)^2 \;T_e(z)\;
    \Lambda(T_e, Z_{ab}) 
  ,
\end{equation}
where $\Lambda(T_e, Z_{ab})$ is the cooling function integrated over
the de-redshifted energy band of observations and $Z_{ab}$ is the
metal abundance of the gas.

On any given LOS, Equations~\ref{E:SZ_1} and \ref{E:THBR_1}, and the
emission weighted X-ray temperature, $\bar{T_e}$ (from spectroscopy), 
provide three independent integral constraints for the two functions, 
$n_e({\bf r}_{LOS})$ and $T_e({\bf r}_{LOS})$, although these 
functions cannot be determined uniquely. If, instead, one assumes 
parameterized functional forms for $n_e({\bf r})$ and $T_e({\bf r})$, 
these three equations can be used to constrain the controlling
parameters. One important parameter that can be found from fitting the
data is a characteristic LOS physical size of the system,
$R_{char}^{LOS}$, and the angular diameter distance of the cluster can
then be determined if this size is compared with the corresponding
angular size, $\theta_{char}^{LOS}$. However, we can measure only the
apparent characteristic angular size, $\theta_{char}^{SKY}$ of the
system as it appears projected into the plane of the sky (with
corresponding physical size $R_{char}^{SKY}$). The main difficulty in
the SZ/X method is the de-projection of the cluster from its 2-D
images, that is the problem of finding the value of
$\theta_{char}^{LOS}$ from the measured value
$\theta_{char}^{SKY}$. If this can be done, one can determine the
angular diameter distance as
\begin{equation} \label{E:D_A1}
  D_A =  { R_{char}^{LOS} \over \theta_{char}^{LOS} } ,
\end{equation}

If one assumes spherical symmetry, $\theta_{char}^{SKY} =
\theta_{char}^{LOS}$, and a de-projection is possible without further
assumptions about the functional forms of $n_e({\bf r})$ and $T_e({\bf
r})$ \citep{SilkWhit78}. This method, however, needs high
signal/noise and high angular resolution SZ {\it and} X-ray images of
the cluster, and thus the analytic de-projection has not been used so far. 
Most commonly, the distribution of gas in clusters is described by
the spherical isothermal beta model \citep{CavaliereF76}.
This model seems to give a good fit to X-ray data \citep{Sara88},
and thus it is usually assumed that the IC gas follows a
spherical isothermal $\beta$ model, where the IC gas is isothermal
and in hydrostatic equilibrium in the gravitational potential of the
cluster. The electron concentration in the cluster atmosphere is then
$n_e(r) \propto (1+(r/r_c)^2)^{-3 \beta/2}$, 
where $r_c$ is the core radius (a suitable characteristic size of the
cluster). The shape parameter $\beta$ describes how the kinetic energy
is distributed between the galaxies and IC gas. Typically 
$r_c$ = 0.1 h$_{100}^{-1}$ Mpc, and $\beta \approx 2/3$ \citep{Sara88}. 
In practice, the IC gas temperature is determined from X-ray
spectroscopy, the core radius projected in the plane of the sky,
$\theta_c^{FIT}$, and the shape parameter $\beta^{FIT}$ are determined
from fitting the $\beta$ model to a high resolution X-ray
image. Finally equations~\ref{E:SZ_1} and \ref{E:THBR_1} are used to
determine the angular diameter distance using Equation~\ref{E:D_A1}
\begin{equation} \label{E:1}
  D_A = { r_c^{FIT} \over \theta_c^{FIT} } = {r_c^{LOS} \over
        \theta_c^{FIT} } 
  ,
\end{equation}
which is crucially dependent on the assumption of spherical symmetry:
$r_c^{FIT} = r_c^{LOS}$. 

If we adopt the beta model representation, we can write the angular
diameter distance as a function of measurable quantities for a cluster as 
\begin{equation} \label{E:D_A_prop}
  D_A \propto { 1 \over (1 + z)^4 } 
              { (\Delta T_0)^2 \over S_{X0} (N_H) } 
              { \Lambda_e(T_e, Z_{ab}) \over T_e^2 } 
              { F(\beta) \over \; g^2(\nu)\; \theta_c } 
  ,
\end{equation}
where $\Delta T_0$ and $S_{X0}$ are the central SZ effect and X-ray
surface brightness (un-absorbed), $N_H$, is the absorbing column
density, and $F(\beta)$ is a known function of $\beta$ (or other
parameters if a different description of the structure of the cluster
is used). Some of the measured (or estimated) quantities on the RHS of
this equation are highly correlated, so a careful error analysis is
needed: the errors on individual quantities cannot be added in
quadrature. No bootstrapping is required in using Eq.~\ref{E:D_A_prop}
(it is a direct method, with no need for using the usual cosmic
distance ladder) as was realized by Sunyaev \& Zel'dovich (1980), but
the strong assumptions of spherical symmetry and isothermality are
only crude approximations to the real situation.

It has been pointed out recently that relativistic effects become
important for high temperature clusters \citep{Reph95}. Relativistic
corrections to the inverse Compton scattering have been calculated and
used to interpret observations in terms of the Hubble constant
(Rephaeli 1995; Birkinshaw 1999; Birkinshaw \& Hughes 1998; Holzapfel
et al. 1997, and references therein). Relativistic corrections to the
thermal bremsstrahlung cooling function are also needed
\citep{HughBirk98,RephYank97}. These corrections are in the order of
3$\%$ in $D_A$ for high temperature ($T_e >$ 10 keV) cluster.  
In this paper we assume that relativistic corrections have been
made and we will not discuss their effect in detail.

% % % % % % % % % % % % % % % % % % % % % % % % % % % % % % % % % % % % % % % % % % % % % % % % %

\section{Error budget in the determination of the angular diameter distance}
\label{s:errors}

Errors in the determination of the angular diameter distance may be
cast into two major categories: errors from measurements, and errors
from theoretical modeling. Measurement errors can be statistical or
systematic, while errors from theoretical modeling are systematic by
nature. The systematic errors can be further classified as ``random''
or ``non-random'' depending on whether they average out or not for a
statistical sample of clusters. In what follows we therefore use the
expressions ``random modeling errors'' when discussing errors which
introduce only scatter in the distance determination when averaged
over an unbiased sample of clusters, and ``non-random modeling
errors'' when discussing errors which introduce a bias in the distance
determinations even for an unbiased sample of clusters.

In this Section we summarize the statistical and systematic errors
associated with the SZ and X-ray observations, following the
discussions in reports of the latest interferometric observations
(from BIMA,  Reese et al. 2000; and from the Ryle Telescope [RT],
Grainge et al. 1999), and in review articles (Birkinshaw 1999). We
also use some additional references to compile a detailed list of
error sources.

\subsection{Errors from measurements}

Statistical errors from measurements in the angular diameter distance
are dominated by counting statistics in X-ray images and spectra, and
by Gaussian measurement uncertainties in the SZ measurements. These
statistical errors propagate to errors in $\theta_c$ and $\beta$
through fitting for the spatial distribution, and to errors in $T_e$,
$\Lambda_e$, $Z_{ab}$, and $N_H$ through fitting the X-ray spectrum. 
A simultaneous fit of a $\beta$ model on interferometric SZ and
X-ray images determines $\theta_c$, $\beta$, $S_{X0}$ and $\Delta T_0$
(e.g. Reese et al. 2000), but causes the errors in these parameters to be
strongly correlated.

The most important error sources in a determination of the angular
diameter distance are the error in $S_{X0}$ and the error in $T_e$
(Eq.~\ref{E:D_A_prop}). The electron temperature enters the estimate
of $D_A$ roughly as $T_e^{-2}$, since for most X-ray observations to
date the X-ray emissivity, $\Lambda_e(T_e,Z_{ab})$ is a slowly-varying
function of $T_e$. Other uncertainties, from the dependence of
$\Lambda_e$ on the abundance of metals in the ICM and the absorbing
column on the line of sight, are less important, and we can also
neglect the contribution to the error in the angular diameter distance
from the redshift.

The error in the spatial fit is dominated by the uncertainty in the
central SZ decrement, $\Delta T_0$, and central X-ray surface
brightness, $S_{X0}$. Both errors are about 8-10$\%$ in the data
obtained using BIMA and ROSAT. The total uncertainty in a $D_A$
estimate from spatial fitting to X-ray imaging and SZ interferometric
measurements (with BIMA or the RT) is about 14-18$\%$. We can expect a
dramatic improvement in the accuracy of $S_{X0}$ in the near future
because of the larger collecting area and higher angular resolution of
XMM and Chandra, and because the improved imaging will also allow a
better choice of models for the ICM. A substantial improvement in the
central SZ effect is also likely, as the first generation of dedicated
SZ interferometers becomes available. 
The overall statistical error in these parameters should drop to 
about 4-5$\%$.

The statistical errors in measurements of electron temperature
based on ASCA and ROSAT observations 
are up to about 15-20$\%$ \citep{Rees00,Grainet99}.
With today's technology (Chandra and XMM), a 5$\%$
statistical error in the emission-weighted average electron
temperature of a well-defined region of a high redshift cluster should be
straightforward \citep{Daviet01,Arnaet01,Peteet01}.

The principal identified systematic errors in the measurements arise
from the absolute calibration of the radio and X-ray observations.
Errors in the effective area of the ROSAT PSPC and HRI introduce an
error of about 10$\%$ in $D_A$. This error is greatly reduced (to
1--2$\%$) for the instruments on Chandra and XMM-Newton. The absolute
calibration error of radio interferometers is good to 4-5$\%$, but
could be improved to about 1$\%$ through extensive observations of
point sources, tied to the planetary flux density/brightness
temperature scale \citep{Birk99}.

Ground-based interferometric observations are also subject to
systematic errors due to the removal of background (and cluster) radio
sources, which may be imperfect if the sources are variable or have
significant angular extent. Imperfect calibration of the
phase and amplitude of the detector system may also cause errors near
brighter contaminating sources. Fortunately interferometers are 
insensitive to large scale gradients in emission from the ground and
atmosphere, and so this source of systematic offset signals from
single-dish data is largely removed in interferometric work (e.g., 
Carlstrom et al. 2001).

\subsection{Errors from modeling}

The modeling of the structure of the ICM is an important part of
determining the distance to a cluster, since it is the model that
makes it possible to connect the LOS size of a cluster to its
apparent angular size projected on the plane of the sky. Following
most treatments in the literature, we use isothermal spherical $\beta$
models to describe the IC gas distribution, and in this Section
discuss the errors introduced by deviations from this model.  As
before, we can identify random and non-random systematic modeling
errors.

Significant random modeling errors arise from the {\it asphericity} of
the intra-cluster gas, the {\it peculiar velocity} of the cluster, and
{\it primordial CMB fluctuations}. Serious non-random modeling errors
can be expected from {\it non-isothermality}, {\it cooling flows},
{\it clumping}, {\it merging}, and {\it finite extent} of the
cluster. Other issues that may be important are {\it radio point
sources} and {\it gravitational lensing}. We briefly discuss random
modeling errors from resolved radio halos in clusters because of their
contribution to the measured SZ signal (through synchrotron emission)
and to the X-ray signal (through inverse Compton emission), 
the effect of diffuse free-free emission from cool gas (for example in
spiral galaxies), and finite optical depth effects.

\clearpage

\subsubsection{Random modeling errors}

{\it Asphericity} is one of the most important source of systematic
errors in the determination of the distance to clusters when using a
spherically symmetric isothermal $\beta$ model. If the cluster is not
spherical, the assumption that $r_c^{FIT} = r_c^{LOS}$ is invalid. 
We estimate the extent of this problem by approximating the true
structures of clusters as oblate or prolate ellipsoids (or, more
generally, as triaxial ellipsoids), while retaining the assumption
that the distribution is described by an isothermal $\beta$ model with
constant $\beta$. If the symmetry axis is in the LOS, we would assume
$r_c^{FIT} = r_c^{LOS}$, and thus overestimate (underestimate) $D_A$ for
prolate (oblate) clusters. \cite{Birk99} finds that prolate or
oblate clusters with symmetry axis aligned in the LOS, if assumed
spherical, will have a fractional error of  
\begin{equation} \label{E:2}
  {\delta D_A \over D_A} = { a - b \over b }
,
\end{equation}
where $a (= r_c^{LOS})$ is the core radius in the LOS and $b (=
r_c^{FIT})$ is the core radius in the plane of the sky ($a > b$ for a
prolate distribution). Clusters often show ellipticity at the level,
in projection, of $a/b = 1.25$ \citep{Mohret95}. If the axis of
symmetry is not in the LOS, the error is smaller, therefore the
fractional error in $D_A$ should be $< 25 \%$. \cite{HughBirk98}'s 
analysis of CL0016+16 showed that oblate or prolate
distributions may cause less than 8$\%$ error in $D_A$ if the structure
of the ICM is analyzed as spherical. \cite{Grainet99}'s work implies
about a 14$\%$ error in $D_A$ for Abell 1413.

N-body simulations can be used to understand the details of physics
of the cluster geometry, and quantify the deviations from the
spherical distribution. 
\cite{Inaget95}'s numerical simulations show that 
asphericity causes an error of up to 15$\%$ in $D_A$. 
Cluster merging simulations of \cite{Roetet97} show that 
small off axis merging in general causes prolate distributions, while
oblate distributions may be caused by large off axis merging, causing
typically less than 20$\%$ error in $D_A$.
Observationally, \cite{BasPM00} found that 
prolate spheroid models fit the APM cluster data better than oblate spheroids.
\cite{Sulk99} studied a statistical sample of clusters assuming a
triaxial beta model density distribution. 
Sulkanen's results indicate that the distance scale obtained assuming
a spherical distribution is within 14$\%$ of its true value (at 99.7\%
confidence) based on a sample of 25 clusters with triaxial axes
consistent with observations (cf. also Puy et al. 2000).

In general, oblate or prolate clusters will have their axis randomly
distributed in the sky. \cite{Zaroet98} studied general
de-projections assuming that clusters have axially-symmetric density  
distributions with an arbitrary orientation of the symmetry axis.
They found that using SZ and X-ray images one can determine only a
combination of distance and inclination angle. Another image (weak
lensing for example) is necessary to decouple these two parameters and
determine the distance separately. If only X-ray and SZ data are
available, the error that asphericity introduces into the $D_A$
determination for any single cluster cannot be reduced below about
15$\%$. However, the combination of X-ray, SZ and weak lensing data
together with modeling of the equilibrium of the hot gas, 
should allow us to reduce the error to around 5$\%$. The planned weak
lensing surveys should make the required lensing data available in the
near future. Alternatively, galaxy velocity distributions from optical 
observations can be used with numerical simulations to determine physical 
parameters of individual clusters, as was done by \cite{Gomeet00} and 
\cite{Roetet97} (see the discussion of merging, below).

{\it Peculiar velocities} of clusters introduce enhanced
(approaching) or decreased (receding cluster) SZ measurements because
of the kinematic SZ effect (Sunyaev \& Zel'dovich 1980; for a
derivation cf. Birkinshaw 1999). In CDM models, cluster peculiar
velocities are 400-500~$\rm km\,s^{-1}$ \citep{Col00,Uedet93}. This would
introduce only a few percent error in $D_A$. However, some
observations suggest larger peculiar velocities, 1000~$\rm km\,s^{-1}$
\citep{BahcSone83,LauePost94}. If these large
velocities are real, the kinematic SZ effect may cause an error of up
to 25$\%$ in $D_A$. Fortunately one can separate the static and
kinematic SZ effects based on their different frequency dependence or,
equivalently, the peculiar velocities can be determined by
measuring the cross-over frequency of the total SZ effect (e.g., 
Molnar \& Birkinshaw 1999). Peculiar velocities also introduce a
small bias in the redshift determination of the cluster. If clusters
are selected based on their SZ signal, this effect would cause a
biased sample of clusters, and a systematic overestimate of $D_A$
would result.

{\it Primordial CMB fluctuations} introduce systematic effects in the
distance determinations with their positive or negative contributions
to the microwave decrement misinterpreted as SZ signal. The amplitude
of systematic errors introduced by CMB fluctuations is a strong
function of the observation strategy. The CMB fluctuations are reduced at
arcminute angular scales compared to their degree-scale amplitudes,
but still introduce a scatter of about 10$\%$ in the distance
determinations \citep{Cen_98}. At smaller scales (high $\ell$s) the
power in CMB fluctuations becomes negligible. A further reduction in
the level of this error can be achieved using spectral separation of
the thermal SZ effect from the primordial fluctuations (and the
kinematic SZ effect).
Gravitational lensing transfers power from large scale primordial 
fluctuations to small scale fluctuations \citep{MetcSilk98,Sel96}. As a 
result, if the CMB  is not separated from the SZ effect, it would give 
larger, but a symmetric scatter in $D_A$, or a fractional error of about 
8$\%$ \citep{Cen_98}. 
Note, however, that the kinematic SZ effect and primordial CMB 
fluctuations have the same frequency dependence, thus they can 
not be separated from each other based on their frequency signature. 
Fortunately both effects can be separated out from the static SZ effect 
simultaneously, based on their different frequency dependence.

The errors in the cosmological parameters caused by random modeling
errors  (asphericity, cluster peculiar velocities, and primordial
fluctuations, etc...) can be reduced using a properly-selected sample
of clusters. This sample must avoid selection biases that themselves
introduce systematic errors. As was emphasized by \citep{Birket991},
clusters should not be selected based on their SZ or X-ray central
brightnesses, since such a selection would produce a sample containing
an excess of clusters with prolate geometry, high positive peculiar
velocity in the LOS, and contamination from negative CMB fluctuations
(if the Rayleigh-Jeans frequency band is used), and result in a biased
$D_A$. X-ray selected clusters with a flux limit well above the
detection limit might be used as in \cite{Masoet01} and
\cite{Joneet00}. Alternatively, if there are multi-frequency
measurements, the best solution would be to separate the kinematic SZ
effect and the primordial fluctuations from the static SZ effect, and
use the total static SZ effect flux density as a selection criterion.

% % % % % % % % % % % % % % % % % % % % % % % % % % % % % % % % % % % % % % % % % % % % % % % % %
\subsubsection{Non-random modeling errors}

{\it Non-isothermality} is one of the major sources of systematic
error in the determination of $D_A$.

The isothermal assumption for the intra-cluster gas is clearly an
approximation. Even if the central region of cluster is virialized
and isothermal, the outer regions will be subject to shocks from
merging and gas in-falling from filaments. Merging with massive
clusters will change the temperature relative to the single-component
cluster virialized value, even in the core region. Observations show
that the central regions are nearly isothermal, but thermal
substructures have also been found in clusters \citep{Sara88}. 
The effect of temperature variations on the distance estimate depends 
on the instrument and observing technique used. SZ measurements 
are insensitive to temperature variations in the cluster if the 
projected pressure profile is unchanged. Non-spatially resolved 
X-ray measurements determine emission weighted temperatures, and so are
sensitive to the thermal structure of the central region of clusters. 
Thus the temperature deduced from X-ray measurements is well suited to
comparison with SZ measurements by radio interferometers, 
which also most sensitive to the central region and have less response
to the outer parts of the cluster
(for example, assuming 2$\sigma$ detection, BIMA is sensitive out 
to about 3$r_c$, and about 85$\%$ of the observable X-ray flux
of a typical cluster, using ROSAT PSPC, is within 3$r_c$).

\cite{BirkHugh94} and \cite{Holzet97} analyzed Abell 2218 and Abell
2163 respectively, assuming an isothermal $\beta$ model and a model
with falling temperature with radius. Their results show that $D_A$
may be overestimated by 20-30$\%$ if non-isothermal distributions are
assumed to be isothermal.  Numerical simulations show similar errors
due to non-isothermality: thus \cite{Inaget95}'s simulations lead us
to conclude that an overestimate of 25$\%$ in $D_A$, mostly because of
the overestimated SZ amplitude, may result from assuming isothermality
when the temperature is lower in the outer regions. Simulations of
merging clusters by \cite{Roetet97} obtained a similar result, that an
overestimate of 10-30$\%$ may result in $D_A$ from non-isothermality,
with the range depending on the projection geometry. Note, however,
that these results are based on single-dish measurements: for
interferometric observations these effects are usually smaller.

It is not easy to correct for non-isothermality. The SZ effect is
proportional to $< n_e T_e > R_{char}^{LOS}$, and for an 
accurate calculation of the SZ effect, it is often necessary to have
good information about the temperature out to $R_{vir}$ (about $10 \,
r_c$) or more. It is difficult to carry out spatially resolved
spectroscopy at such large radii because of the low X-ray surface
brightness of the outer regions of clusters ($S_X \propto n_e^2$). 
At present, the best evidence from BeppoSAX data extends $T(r)$
measurements to $(0.5 - 0.75) R_{vir}$ \citep{IrwiBreg00,DeGrMole99}.
The increased sensitivity available with Chandra and XMM
should enable us to use spatially resolved spectroscopy to determine
the temperature profiles of clusters which both resolve the inner
cooling flow region and collect enough photons to measure useful
temperatures in the clusters' outer regions (Schmidt, Allen \& Fabian
2001; Tamura et al. 2001).

Unfortunately the hydrostatic equilibrium assumption of the $\beta$
model breaks down in the central parts of clusters with high enough
density for cooling to be important. In these cases the ICM will
radiate via thermal bremsstrahlung and line emission (with the balance
depending on the temperature), and develop a pressure gradient and a
sub-sonic inflow, a so-called {\it cooling flow} (see review of Fabian
1994). The increased central density at the core of
the cluster leads to an increased level of X-ray emission, which is
often used as an indicator of the presence of a cooling flow region.

Phenomenological models have been developed and numerical simulations
have been performed to study cooling flows (White \& Sarazin 1987;
Rizza et al. 2000; Fabian 1994, and references therein). Majumdar and
Nath (2000) estimated the effect of cooling flows on the determination
of the Hubble constant. They found that only at the very center of the
cooling flow (within the sonic radius) will the X-ray luminosity 
drop because of the decreased temperature of the ICM. They show,
further, that there is a gradual increase in pressure towards the
center of the cooling flow. Outside this region, the approximation of 
hydrostatic equilibrium profile remains good, while within there will
be an SZ effect excess. Thus Majumdar \& Nath (2000)'s results
indicate that one should exclude 80$\%$ of the cooling flow region to
reduce the error in $D_A$ to below 10$\%$.
However, this may be an upper limit on the error in $D_A$, since the 
calculation ignored the effect of the cooling flow on the structure 
fitting. Fig.~3 and 4 in Reese et al (2000) suggest that a compensatory 
error occurs here, and this lowers the error in $D_A$.

A further effect from cooling flows, pointed out by Schlickeiser
(1991), is that the build-up of cold gas at the center of cooling
flows might lead to significant free-free emission in the radio band,
which would reduce the SZ signal. This works in the opposite sense to
the pressure effect, but provides another reason for excluding the
central parts of SZ images of cooling flow clusters from the SZ/X
distance scale analysis.

Excluding cooling flow clusters completely from SZ/X distance
scale studies would be the most complete solution to the problem that
they pose. However, most clusters close to hydrostatic equilibrium
(where the underlying physics required to model the ICM is most
straightforward) possess cooling flows. Modeling cooling flows, and
excluding their centers, is therefore necessary to build up large
samples for Hubble constant work.

High-redshift cooling flow clusters can be recognized by using their
X-ray spectra even if we cannot resolve the central region, since
cooling flow clusters have emission weighted metallicity 1.8 times
higher than non-cooling flow clusters \citep{AlleFabi98}.
Mohr, Mathiesen \& Evrard (1999) analyzed nearby cluster of galaxies
using ROSAT PSPC data, modeling cooling flow regions where necessary,
and found that fitting an isothermal $\beta$ model to cooling flow
clusters will underestimate both $r_c$ and $\beta$, and therefore
produce a poor fit even outside the cooling flow. They identified
cooling flow clusters based on two criteria: 1, non-random residuals
consistent with a central emission excess; and 2, relaxed cluster with
no asphericity or substructure. They concluded that a double $\beta$
model fit gives an unbiased estimate of $D_A$.
Cooling flow contamination in the determination of the average Hubble 
constant using the SZ/X ray method can be recognized by searching for 
a dependence of individual Hubble constants (determined for each 
cluster) on IC gas metallicity.

{\it Clumping} in the IC gas is potentially one of the most important
systematic effects. It is well known that radio halos, hot bubbles
from supernova eruptions, cold condensed gas in galaxies,
etc. constitute a level of clumpiness in the ICM. The important
question is whether the effect of clumping is important in the SZ/X
method of measuring $D_A$.

As a first approximation, clumping will enhance the X-ray surface
brightness since $S_X \propto n_e^2$, and does not change $\Delta
T_{SZ}$, since $\Delta T_{SZ}$ is proportional to the line-of-sight
averaged pressure and clumps should be in pressure equilibrium with
their surroundings if they are to be long-lived (assuming no
substantial magnetic fields exist). As a consequence, $D_A$ is
underestimated. However, the emission-weighted temperature of the
cluster measured by X-ray spectroscopy will also decrease. This reduces
the systematic effects of clumping to only a few percent error in $D_A$.
Birkinshaw, Hughes \& Arnaud (1991) studied the effects of isobaric
clumping of the intra-cluster gas in Abell~665. The fractional error
in $D_A$ from clumping is 
\begin{equation} \label{E:3}
  {\delta D_A \over D_A} = { < n_e >^2 <\Lambda (T_e) > \over  < n_e^2
   \Lambda (T_e)> } - 1 
  ,
\end{equation}
where the angle brackets imply averages over regions larger than the
scale of the clumping. For isothermal clumps $\Lambda_e$ factors out.
Then since $< \! n_e^2 \!> /< \! n_e \!>^2$ is always greater or equal
to 1, clumping will always cause an underestimate of $D_A$. Birkinshaw
et al. (1991) found $< \! n_e^2 \! >^2 / < \! n_e \! >^2 < 3$ or
so. Holzapfel et al. (1997) studied the effects of isobaric clumping
on Abell 2163. Assuming cold clumps they estimate that error from
clumping could lead to an overestimate of about 10$\%$ in $D_A$.

Inagaki et al. (1995) find, from their $\Omega_m = 1$ numerical
simulations, that $D_A$ will be underestimated by about 15$\%$ due to
clumping. However there is evidence that lower matter density  
models produce less clumpy structures, so that if our Universe has 
$\Omega_m < 1$, Inagaki et al. will have overestimated the effect of
clumping.

Self-consistent modeling of small scale clumping is difficult because
many physical processes contribute to its creation and destruction.
\cite{GunnThom96} used phenomenological multi-phase models to study
X-ray emission from clumpy IC gas. Their isobaric model implies a
fractional error in the X-ray central surface brightness 
\begin{equation} \label{E:1}
  {\delta S_X(0) \over S_{x0} }
     \propto 
     {< n_e > <n_e^{1-\alpha}>^{\alpha/2}  \over
      <n_e^{2-\alpha}>^{(1+\alpha)/2} } 
  ,
\end{equation}
where $\alpha$ is the emissivity exponent. This would cause about a
10$\%$ error in $D_A$.

\cite{Nagaet00} discussed biases in the Hubble constant determination
from a multi-phase, spherically-symmetric, intra-cluster medium with
isobaric clumping of variance
\begin{equation} \label{E:1}
  \sigma^2 = {\sigma_c^2 \over (1 + (r/r_c)^2 )^{\epsilon} }
,
\end{equation}
where $\sigma_c$ and $\epsilon$ are free parameters describing the
strength and radial dependence of the variance. They assumed a
log-normal distribution for the gas density phase distribution, which
was motivated by simplicity and the effect of non-linear gravitational
growth of initially Gaussian density fluctuations (Cole, Fisher and
Weinberg 1994). They assumed that the multi-phase model has the same
emission weighted temperature and the X-ray emission profile as a
fiducial single phase model. Based on their results, the error in the
distance from an incorrect assumption of a single-phase ICM is
\begin{equation} \label{E:1}
  {\delta D_A \over D_A} = 2 \exp [ {(1-\alpha)(2-\alpha) \over 4}
  \sigma_c^2 ] - 1 
  ,
\end{equation}
where $\alpha$ is again the power-law exponent in the emissivity
function. We can conclude that clumping may cause a $5 - 20 \%$ error
in $D_A$.

Spatially-resolved X-ray spectroscopy, as can be performed using
Chandra and XMM, will help to estimate the clumpiness of the IC gas.
Emission lines between 0.5 and 1.5~keV originating in cool regions,
such as the Fe L-shell lines, H- and He-like lines
from N, O, Ne, Mg, will provide a strong test on multi-phase models.
Unfortunately this method will not constrain all types of clumpiness.
\cite{HughBirk98}, and \cite{Maso99} suggest that, since SZ/X Hubble
constant determinations are not very different from the results of
other methods, clumping introduces less than a 10$\%$ error. 
However, it is possible that clumping itself introduces more bias, but
that this bias is balanced by other effects with opposing biases.

\cite{Roetet97} studied systematic errors in the Hubble constant from
cluster {\it merging}. Merging can lead to the formation of shocks
that compress and heat the IC gas. After a merger, the gas will settle
into the prolate or oblate potential well defined by the dark matter
distribution. As already discussed, ellipticity introduces scatter
rather than bias in $D_A$. However, shock-heated IC gas will have
different SZ and X-ray properties than gas in hydrostatic equilibrium,
and its presence will introduce bias. Based on simulated X-ray images
and temperature maps, Roettinger et al. find that isothermal
assumptions in merging clusters systematically overestimate $D_A$ by
about 15$\%$. Numerical simulations show that density enhancements due
to merging of sub-clusters can result in higher X-ray surface
brightness, change mass estimates, and might, if not recognized, cause
an about 20$\%$ error in distance determination \citep{Mohret99}.

\cite{Roetet97} suggest that clusters at the early stages of merging
should be excluded from distance determinations. Dynamically active
clusters may be recognized from their galaxy velocity
distributions. Clusters with dynamical activity will have a large
$\beta$ discrepancy ($\beta_{fit} \ne \beta_{spect}$, see details in
Sarazin 1988). Anisotropy in  galaxy velocity distribution also
signals dynamical activity. More relaxed systems, like merging
clusters after they reached quasi-equilibrium may be modeled and
included in the distance determination. Simulations help to analyze
individual merging clusters \citep{Roetet95,Gomeet00}: adjustments to
the initial parameters of merging clusters are made until the
resulting merged cluster has the observed SZ and X-ray appearance and
galaxy velocity distribution.

The $\beta$ model gives divergent masses if not truncated at some
finite radius, which is usually taken to be about 10 core radii.
This fact is a sign that at large radii, the beta model can not be correct.
When calculating the distance to a cluster using the isothermal $\beta$ model,
we assume an infinite extent, and therefore the {\it finite extent} of clusters
introduces a systematic bias in $D_A$.
The SZ effect is more sensitive to outer regions than X-ray bremsstrahlung, 
since it is proportional to $n_e$ unlike X-ray bremsstrahlung,
which is prop to  $n_e^2$. In theory, the SZ effect measurements are 
more suitable for studying the outer regions of clusters, and also
are subject of more error caused by the finite extent of clusters.

\cite{Inaget95} discussed the effect of finite cluster sizes. They
find that the ratio between finite truncated and full beta models is
\begin{equation} \label{E:1}
  {\delta D_A \over D_A} = \Biggl[
                             { 1 - { B_q(3\beta/2-1/2, 1/2) \over
                                     B(3\beta/2-1/2, 1/2)}         
                             \over
                               1 -  { B_q(3\beta-1/2, 1/2) \over 
                                      B(3\beta-1/2, 1/2)}         } \Biggr]^2
,
\end{equation}
where $q = 1/(1 + p^2)$, $p = R_{cut}/r_{core}$, $B$ and $B_q$ are the
beta and incomplete beta functions, and assuming no change in
$\theta_c^{FIT}$. \cite{Inaget95}'s results suggest that this effect can 
cause an about 10-20$\%$ underestimation of $D_A$ for typical parameter 
values (cf. also Puy et al. 2000).
By contrast, \cite{BirkHugh94} and \cite{Holzet97}, analyzing Abell~2218
and Abell~2163 respectively, find that finite cluster extent contributes 
only about 6$\%$ and 2$\%$ overestimate error in $D_A$.
Observationally, \cite{SMM00} attempted to find the outer cut off of 
Abell 3571 using RXTE scans across the cluster. 
Abell 3571 seems to extend well beyond its virial radius, however,
this conclusion is not strong due to poor statistics in the data.

Clearly, more observations are needed to find out the extent to which
the $\beta$ model is correct, and how the transition happens from the
cluster to the surrounding regions (dominated by filaments according
to numerical simulations). A falling temperature profile with radius
is a sign of deviations from the $\beta$ model, and may provide the
best tracer of additional structure of this type. We expect errors
from this effect to be reduced greatly when the new spectral
information from Chandra and XMM is available.

One of the most important contaminants of the SZ effect is emission
from {\it radio point sources}. Such emission, especially at the
center of the cluster, will decrease the fitted amplitude of the SZ
decrement in the Rayleigh-Jeans frequency region, and cause an
underestimate in the distance of the cluster. Using the sensitivity of
the instrument, one can estimate the maximum flux density of unresolved  
point sources that can contribute to this error: for BIMA and the RT,
the systematic errors are in the 10-15$\%$ range. The level of radio
source confusion varies with the observing technique
used. Multi-baseline interferometric observations have the most
favorable confusion level. A deep survey, carried out at different
frequencies or interferometer baselines, would help to find point
sources up to a limit when the error due to unresolved point sources
would be negligible relative to other errors. Interferometric
measurements also have the advantage over single-dish observations of
being able to monitor the 
brightnesses of (potentially-variable) point sources while
simultaneously measuring the SZ effect.

As \cite{LoebRefr97} pointed out, systematic effects in the
determination of the baseline for the SZ effect arise from {\it
gravitational lensing} by the cluster gravitational potential.  
The brightnesses of point sources are enhanced by lensing, which
brings them above the detection threshold, and thus they are removed
from the field. This over-removal of point sources from the background
lowers the background flux relative to a control field, and thus leads
to an overestimate of the SZ signal, and an underestimate of about
10$\%$ of $D_A$. However, this effect is important only at frequencies
less than about 30 GHz, and its presence can be tested (and corrected
for) using the model cluster mass distribution which can be obtained
by conventional analyses of the X-ray data for each cluster
(cf. also Blain 1998).

Other, probably small, effects contributing to the error budget, which 
should be checked and treated individually when necessary are:
contributions in the radio band from synchrotron emission from resolved
halo (and other) sources in clusters, free-free emission from cool gas
in spirals, free-free emission from radio halos in the X-ray band
\citep{Birk79}, and any contribution from the non-thermal SZ effect. 
As was pointed out by \cite{MolnBirk99}, the Kompaneets equation and 
its relativistic extensions are equivalent to a single scattering
approximation, thus a small effect will arise from ignoring finite
optical depth (multiple scattering). Any non-thermal population of IC
electrons would produce a non-thermal SZ effect. There are some
theoretical arguments for such a population
\citep{Petr01,Blas00,Colaf99,Sara99}, and some observations have
detected the expected excess hard X-ray emission 
(e.g., in the Coma cluster), but such excess X-ray emission is rare
and hence the non-thermal electron population is weak
\citep{Fusc01,MaloBlan01}. 
Clusters with complex morphology should not be used for distance
determinations because of the difficulty in their modeling.
If substructure in high redshift clusters 
(as in RX J1347-1145, cf. Komatsu et al. 2001) is common, one should
check such individual clusters for complex morphology using a high 
resolution instrument.

The {\it cumulative} systematic effect resulting from all these
sources of error is best addressed via numerical simulations.
\cite{Yoshet98} used numerical simulations to estimate the 
systematic error from fitting $\beta$ models to clusters. 
They used a spatially flat fiducial model with 
$\Omega_m$ = 0.3, $\Omega_\Lambda$ = 0.7, and $h$ = 0.7 for their
simulations. They found that the isothermal $\beta$ model describes
clusters well both in SZ and X-ray imaging. They found that, even
though the fitted $\beta$ model parameters were different when fitted
to SZ or X-ray images (due to non-isothermality, non-asphericity, and
clumpiness), the systematic errors in the Hubble constant (and so in
distances) are negligible at low redshifts. At high redshift ($z \sim
1$), about a 20$\%$ overestimate of $D_A$ might occur, mainly due to
non-isothermality and asphericity.
\cite{Inaget95} find that the cumulative effect of non-isothermality 
and asphericity leads to about a $10 - 20\%$ overestimate of $D_A$,
since the overestimate due to non-isothermality is larger than the
underestimate due to clumpiness (though in their simulations 
$\Omega_m = 1$).

\subsection{Summary of the error budget}

The new generation of satellites will allow us to determine more
precisely the temperature profile and spatial structure of the
intra-cluster medium, and thus minimize the systematic errors in $D_A$
that result from temperature variations. When necessary, the spatial
resolution of these new instruments will allow us to model the gas in
each individual cluster beyond the spherical isothermal beta
model. Numerical modeling of individual clusters will help us to
derive their physical parameters. The improved spatial resolution will
also allow us to study clumpiness, which is another important source
of uncertainty. In general, those effects which have different
spectral signature from the static SZ effect, should be separated from
the SZ effect using multi-frequency observations. 

Based on our evaluation of the error budget, we estimate that 
the random error in $D_A$ achievable in the near future using known
techniques, and assuming that lensing measurements will be used to
eliminate errors from cluster asphericity, might be as low as
7$\%$. This estimate is made up by quadrature combination of the
errors, with the dominant terms being from $T_e$ (5$\%$) and 
spatial fitting (5$\%$). A systematic error of 5$\%$ might also be
obtained, although this would be difficult.

Evolution effects are clearly important limitations in determining
cosmological parameters. The advantage of the SZ/X-ray method 
is that, as long as hydrostatic equilibrium and simple geometry hold, 
it should be reliable even if scaling laws (for example 
mass - temperature) evolve. However, a detailed analysis of evolution 
effects is out of the scope of our paper. We restrict ourselves to 
simply studying the effect of an additional systematic error in the 
$D_A$ determination with a linear gradient with redshift.

% % % % % % % % % % % % % % % % % % % % % % % % % % % % % % % % % % % % % % % % % % % % % % % % %
\section{Constraints on Cosmological Parameters}
\label{s:Constraints}

As discussed earlier, in the near future SZ surveys will discover 
hundreds of high and low redshift clusters, and we can expect Chandra
and XMM observations of hundreds of clusters. However, accurate
angular diameter measurements require long SZ and X-ray integrations
and therefore we do not expect all discovered clusters to have accurate 
SZ/X distance measurements. If the SZ/X method is to be used to
measure cosmological parameters, we should select a sample of clusters
that minimize the systematic errors (Section \ref{s:errors}) while
not requiring excessive observing time. As pointed out in \cite{HuteTurn00}, 
the ideal distribution of clusters in redshift space would be a 
superposition of $P$ delta functions in redshift, where $P$ is the number of
cosmological parameters to be determined. In practice, however, in any
narrow redshift band we will not have enough clusters to achieve good
statistics, and objects at $z > 1$ would need excessive integration
times to reach high individual accuracies. A detailed study to
select the optimum redshift distribution of clusters, taking into
account the differing numbers of suitable clusters at different
redshifts and the detailed characteristics of the possible instruments 
and observational strategies is beyond the scope of our paper and more
properly devolves on the groups proposing to construct such
instruments. Here we do a simpler problem, by assuming that distances
with similar accuracies are available for a set of clusters uniformly
distributed in redshift space between $z = 0.01$ and $z = 1$. 
This choice covers almost half of the redshift interval most sensitive to the
cosmological constant ($z = 0.5 - 1.8$), but excludes the more distant
objects for which excessive integration times would be needed.

We carried out simulations to estimate the constraints from the
angular diameter distance-redshift function on the parameter space
defined by $\Omega_m$, $\Omega_\Lambda$, and $h$, and also by
$\Omega_m$, $w$, and $h$ (assuming a spatially flat geometry
$\Omega_\Lambda = 1 - \Omega_m$). We simulated clusters using a
spatially flat fiducial CDM model with $\Omega_m=0.3$,
$\Omega_\Lambda=0.7$, and $h$ = 0.65.  We use the $\chi^2$ statistic
to evaluate errors from our simulations, since we are dealing with
large errors, 50$\%$ in cosmological parameters, and therefore the
assumptions leading to the Fisher matrix formalism are not satisfied.
We choose realizations which have best fit values close to those of
the input fiducial model, and offset by the values of 
$\Delta \chi^2=$ 3.53, 8.02 and 14.2 
appropriate for the number of fitted parameters to find the Gaussian
1, 2, and 3 $\sigma$ error surfaces (defined by probability levels of
68$\%$, 95.4$\%$, and 99.73$\%$).

\begin{figure}
\figurenum{2a}
\centerline{
\plotone{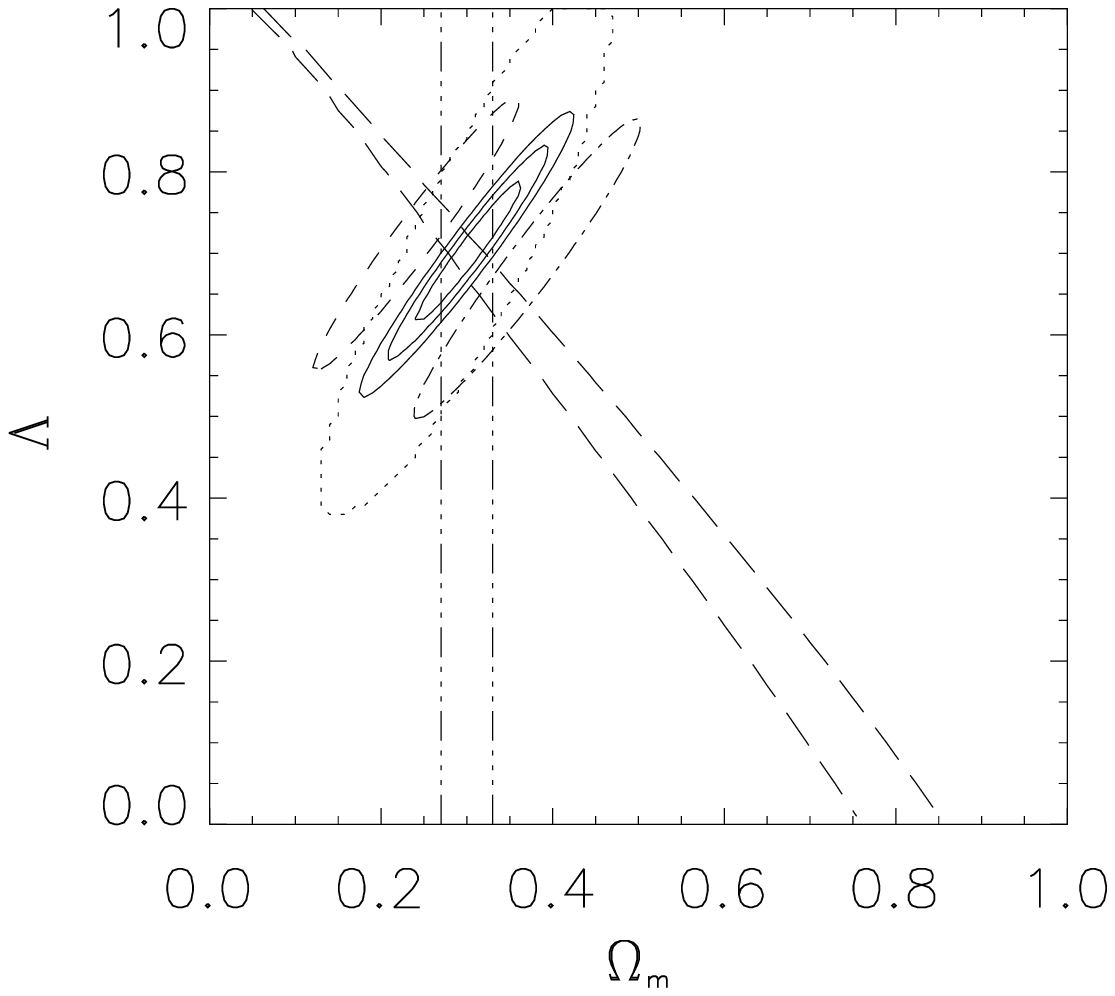}
}
\caption{1, 2 and 3$\sigma$ constraints on cosmological parameters from
simulated angular diameter distance measurements. We used a spatially
flat CDM model as a fiducial model with $\Omega_m=0.3$,
$\Omega_\Lambda=0.7$, and $h$ = 0.65. The figures show results of
simulations using 500 clusters assuming a random error of 4$\%$ in the
angular diameter distance (concentric solid ellipses). 2-dimensional
(2D) projections of the 3D surface of 3$\sigma$ constraints are shown
using dotted lines. 3$\sigma$ constraints assuming an additional 
$\pm 3\%$ systematic error are shown using solid ellipses above and 
below the random ellipses (Figure b and c). Constraints from a
systematic error which grows from 0$\%$ to $\pm 3\%$ at $z=1$ is shown
using short dashed and dash dotted lines. Panels a, b, and c show two
dimensional slices of the three dimensional parameter space defined by
($\Omega_m$, $\Omega_\Lambda$, $h$) passing through the best-fit model
as defined by the minimum of $\chi^2$. Constraints that might be
derived from the location of the first Doppler peak, assuming
$\ell_{peak} = 245 \, \pm 10$, are shown as  long dashed lines, and
define surfaces almost perpendicular to the  error regions from
$D_A(z)$ in the $\Omega_m$ - $h$ plane. Constraints from $\Omega_m
h^2$ = constant, assuming a 10$\%$ error in its determination from
CMB experiments, are shown using dash-dot-dot lines.
\label{F:FIG2}
}
\end{figure}

\begin{figure}
\figurenum{2b}
\centerline{
\plotone{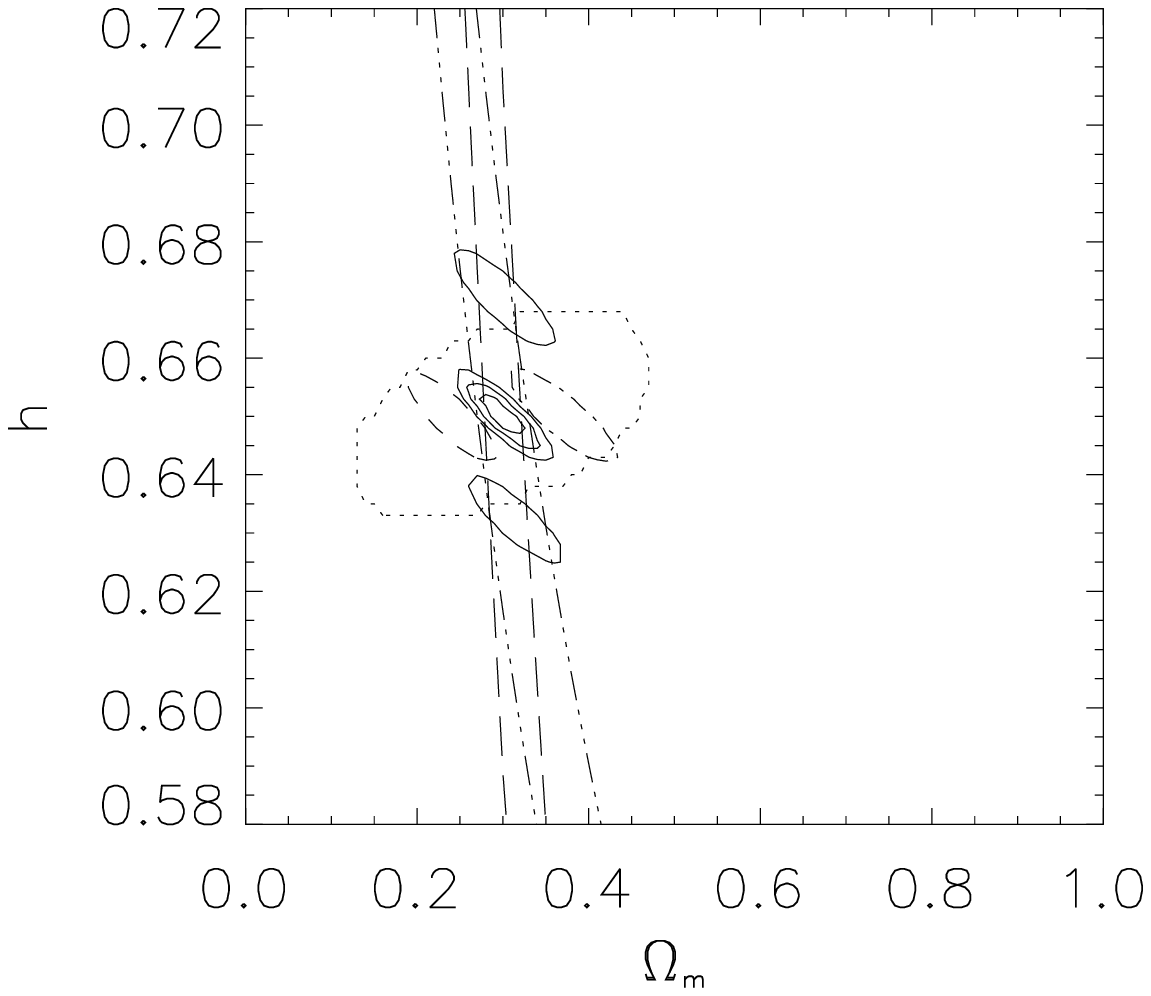}
}
\caption{1, 2 and 3$\sigma$ constraints on cosmological parameters from
simulated angular diameter distance measurements. We used a spatially
flat CDM model as a fiducial model with $\Omega_m=0.3$,
$\Omega_\Lambda=0.7$, and $h$ = 0.65. The figures show results of
simulations using 500 clusters assuming a random error of 4$\%$ in the
angular diameter distance (concentric solid ellipses). 2-dimensional
(2D) projections of the 3D surface of 3$\sigma$ constraints are shown
using dotted lines. 3$\sigma$ constraints assuming an additional 
$\pm 3\%$ systematic error are shown using solid ellipses above and 
below the random ellipses (Figure b and c). Constraints from a
systematic error which grows from 0$\%$ to $\pm 3\%$ at $z=1$ is shown
using short dashed and dash dotted lines. Panels a, b, and c show two
dimensional slices of the three dimensional parameter space defined by
($\Omega_m$, $\Omega_\Lambda$, $h$) passing through the best-fit model
as defined by the minimum of $\chi^2$. Constraints that might be
derived from the location of the first Doppler peak, assuming
$\ell_{peak} = 245 \, \pm 10$, are shown as  long dashed lines, and
define surfaces almost perpendicular to the  error regions from
$D_A(z)$ in the $\Omega_m$ - $h$ plane. Constraints from $\Omega_m
h^2$ = constant, assuming a 10$\%$ error in its determination from
CMB experiments, are shown using dash-dot-dot lines.
\label{F:FIG2b}
}
\end{figure}

\begin{figure}
\figurenum{2c}
\centerline{
\plotone{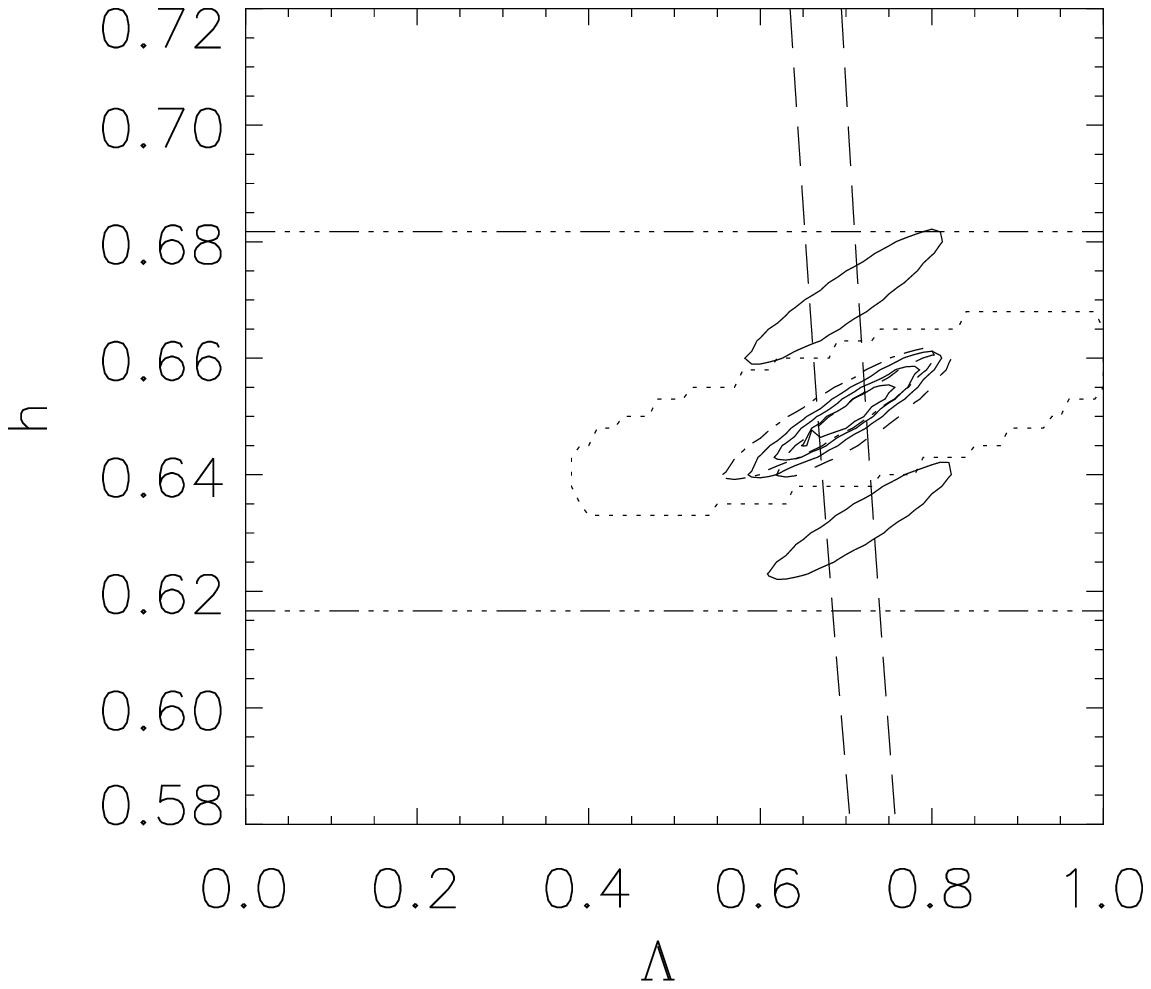}
}
\caption{1, 2 and 3$\sigma$ constraints on cosmological parameters from
simulated angular diameter distance measurements. We used a spatially
flat CDM model as a fiducial model with $\Omega_m=0.3$,
$\Omega_\Lambda=0.7$, and $h$ = 0.65. The figures show results of
simulations using 500 clusters assuming a random error of 4$\%$ in the
angular diameter distance (concentric solid ellipses). 2-dimensional
(2D) projections of the 3D surface of 3$\sigma$ constraints are shown
using dotted lines. 3$\sigma$ constraints assuming an additional 
$\pm 3\%$ systematic error are shown using solid ellipses above and 
below the random ellipses (Figure b and c). Constraints from a
systematic error which grows from 0$\%$ to $\pm 3\%$ at $z=1$ is shown
using short dashed and dash dotted lines. Panels a, b, and c show two
dimensional slices of the three dimensional parameter space defined by
($\Omega_m$, $\Omega_\Lambda$, $h$) passing through the best-fit model
as defined by the minimum of $\chi^2$. Constraints that might be
derived from the location of the first Doppler peak, assuming
$\ell_{peak} = 245 \, \pm 10$, are shown as  long dashed lines, and
define surfaces almost perpendicular to the  error regions from
$D_A(z)$ in the $\Omega_m$ - $h$ plane. Constraints from $\Omega_m
h^2$ = constant, assuming a 10$\%$ error in its determination from
CMB experiments, are shown using dash-dot-dot lines.
\label{F:FIG2c}
}
\end{figure}

As a first realization, we assumed a sample of 500 clusters uniformly
distributed in redshift (so that 250 lie at $z >0.5$) with 4$\%$
random error in $D_A$, which might be achieved if systematic errors
can be tightly controlled (we assume lensing measurements will be used 
to eliminate errors from cluster shape).
We use these results to demonstrate the
ultimate constraints on cosmological parameters that might be 
achievable from the SZ/X-ray method. 
In Figures~\ref{F:FIG2}, b, and c, we show the results of fitting 
cosmological parameters to the observationally-determined angular 
diameter distance function (1, 2, and 3$\sigma$ concentric ellipses,
solid lines, assuming three independent parameters, corresponding to
$\Delta \chi^2$ of 3.53, 8.02 and 14.2 appropriate for three fitted
parameters). 
2-dimensional (2D) projections of the 3D surface of 3$\sigma$ 
constraints are shown using dotted lines.
In the $\Omega_m$ - $\Omega_\Lambda$ plane (Figure~\ref{F:FIG2}) the
error ellipses are elongated roughly along the line $2\, \Omega_m -
\Omega_\Lambda$ = constant. This is similar to the constraint obtained
from the SNe Ia experiment, which measures cosmological parameters via
the predicted apparent magnitude distribution, and uses different
redshift limits. Very low redshift measurements of clusters lead to
error ellipses which are elongated along the line
$\Omega_m - 2\, \Omega_\Lambda$ = constant. As higher redshift objects
are added, this elongated feature rotates counter-clockwise.
From Figure~\ref{F:FIG2c}, which shows constraints in the $h$ -
$\Omega_\Lambda$ plane, we can conclude that with the assumed
accuracy, the cosmological constant is not well constrained.
However, $h$ is well measured even without additional information. 
In general, the dispersion of measurements (1, 2, 3$\sigma$ contours)
is determined by random and not systematic errors.
Systematic errors introduce only bias, shifting the mean of the
measurements away from the expected value in the parameter space
while preserving their dispersion. A $\pm 3\%$ systematic error in the
$D_A$ will not affect the results in the $\Omega_m - \Omega_\Lambda$
plane, but simply shift the error ellipses we obtained from random
errors up or down along the $h$ axis by 0.02 ($3\%$ of the fiducial
value of $h$), since the amplitude of $D_A$ is set by the Hubble
constant (Figures~\ref{F:FIG2}b and c, 3$\sigma$ solid ellipses above
and below the 1, 2, and 3$\sigma$ concentric ellipses corresponding to
random errors, solid lines). 
With the assumed random and systematic errors, the determination of 
$h$ becomes limited by systematic errors.
We also carried out simulations assuming a systematic error with a 
gradient in redshift growing from 0$\%$ at $z=0$
to $\pm 3\%$ at $z=1$ in addition to the assumed 4$\%$ random error 
to study the effect of evolution. 
(Figures~\ref{F:FIG2b} and c, short dashed and dash dotted 3$\sigma$ lines).
The error ellipses from random errors will be shifted in the $\Omega_m$ - 
$\Omega_\Lambda$ plane (Figure~\ref{F:FIG2}).
From Figures~\ref{F:FIG2b} and c we can conclude that the Hubble 
constant will not be affected by this type of systematic errors, 
which follows from the fact that the Hubble constant is constrained by 
the low redshift regime where the assumed systematic errors are small.
$\Omega_m$ and $\Omega_\Lambda$ 
are strongly affected since their relation is determined by redshift 
between 0.5 and 1.8 (cf. Figure~\ref{F:FIG1}), 
where the systematic errors become larger, thus
$\Omega_m$ and $\Omega_\Lambda$ become limited by this type of
systematic errors.
An indication of systematic errors in $D_A$ from evolution might 
be found using an $\Omega_m$ - $h$ plot (Figure~\ref{F:FIG2b}).

\begin{figure}
\figurenum{3a}
\centerline{
\plotone{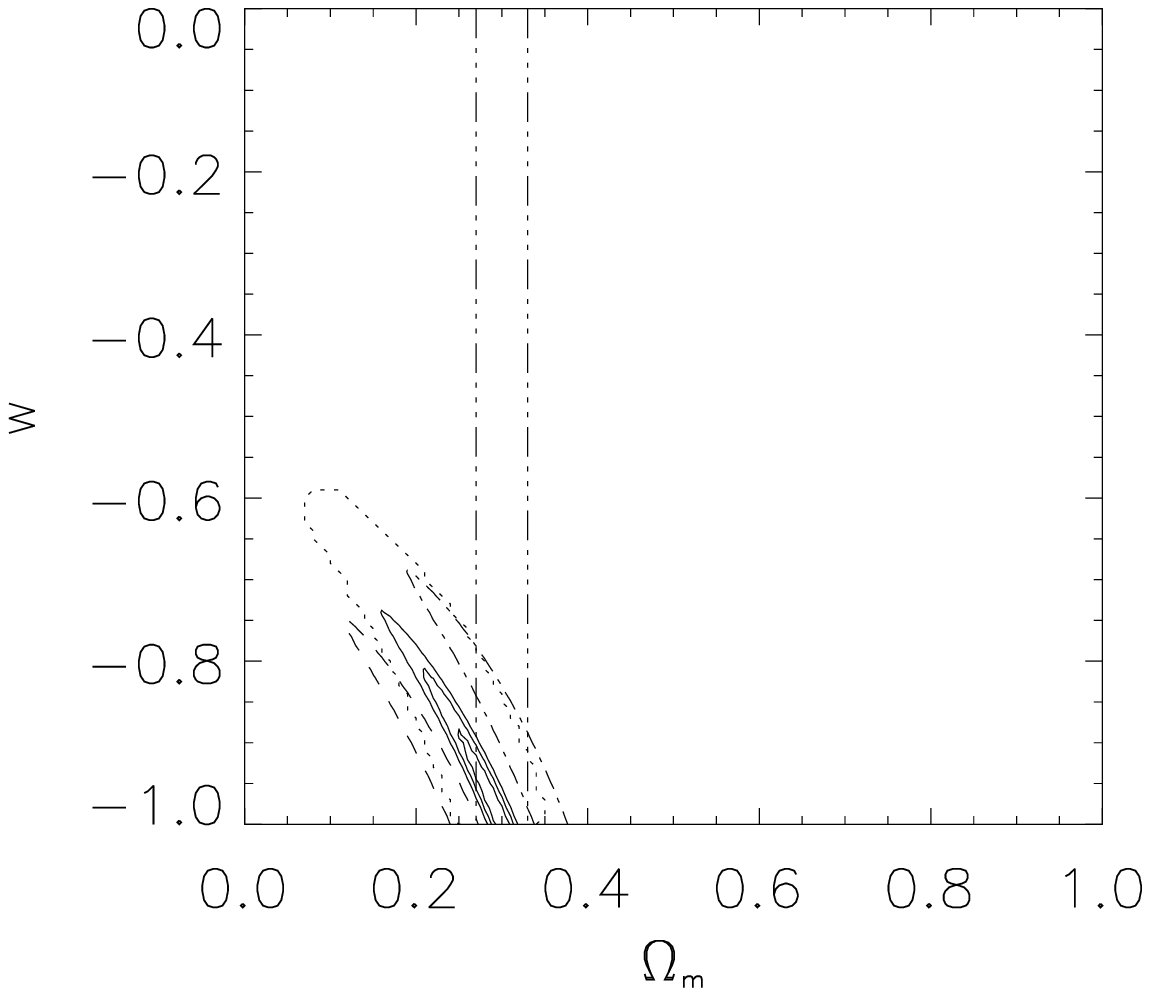}
}
\caption{1, 2 and 3$\sigma$ constraints on cosmological parameters from
simulated angular diameter distance measurements. Contours and lines
have the same meaning as in Figure 2. The figures show results of 
simulations using 500 clusters assuming an error of 4$\%$ in the
angular diameter distance (concentric ellipses, solid lines). 
3$\sigma$ constraints assuming an additional $\pm 3\%$ systematic 
error are shown using solid ellipses above and below the random
ellipses (panels b and c). Panels a, b, and c show two dimensional
slices of the three dimensional parameter space defined by
($\Omega_m$, $w$, and $h$) and passing through the best-fit model
as in Figure 2. Constraints from $\Omega_m h^2$ = constant,
assuming a 10$\%$ error in its determination from CMB experiments, 
are shown using dash-dot-dot lines.
\label{F:FIG3}
}
\end{figure}

\begin{figure}
\figurenum{3b}
\centerline{
\plotone{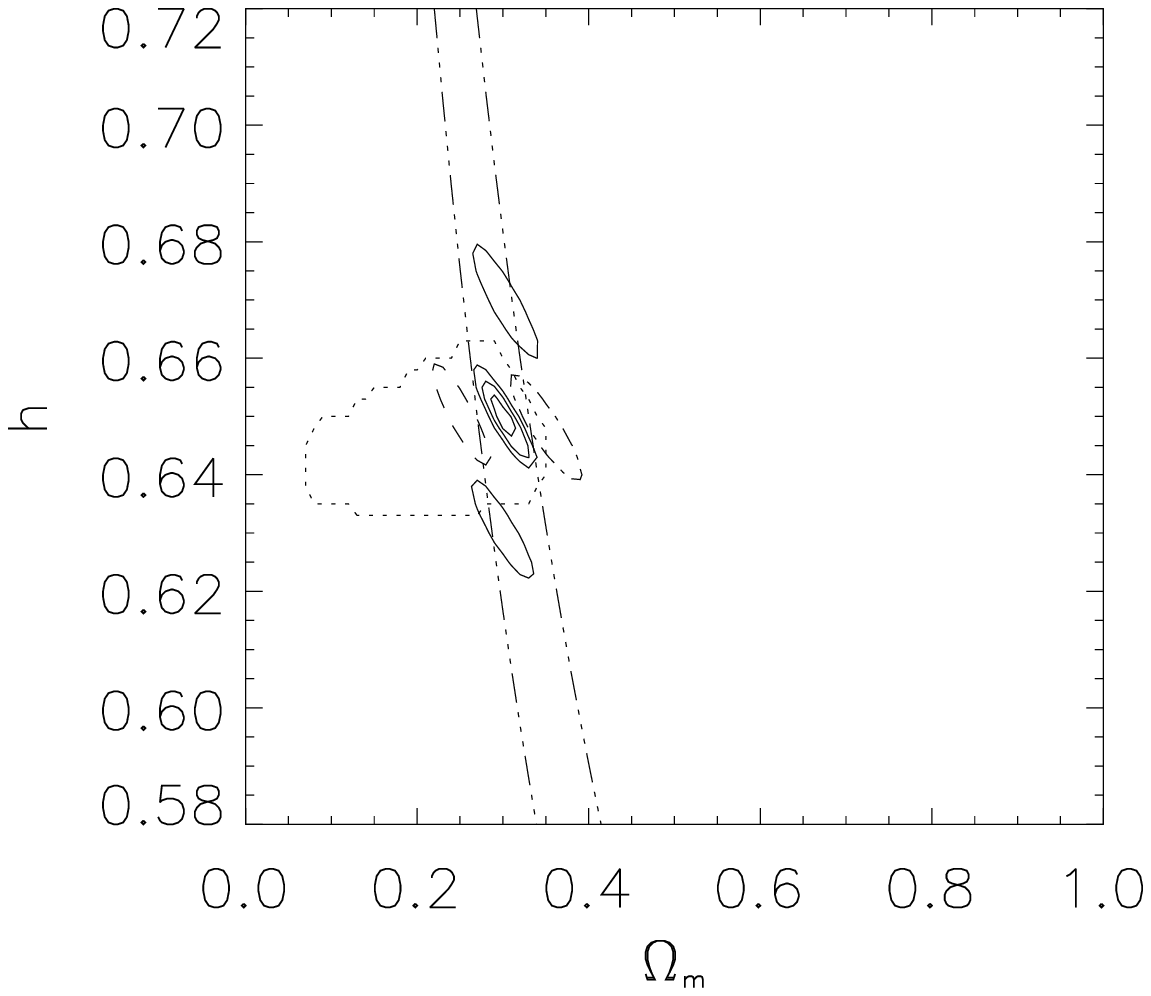}
}
\caption{1, 2 and 3$\sigma$ constraints on cosmological parameters from
simulated angular diameter distance measurements. Contours and lines
have the same meaning as in Figure 2. The figures show results of 
simulations using 500 clusters assuming an error of 4$\%$ in the
angular diameter distance (concentric ellipses, solid lines). 
3$\sigma$ constraints assuming an additional $\pm 3\%$ systematic 
error are shown using solid ellipses above and below the random
ellipses (panels b and c). Panels a, b, and c show two dimensional
slices of the three dimensional parameter space defined by
($\Omega_m$, $w$, and $h$) and passing through the best-fit model
as in Figure 2. Constraints from $\Omega_m h^2$ = constant,
assuming a 10$\%$ error in its determination from CMB experiments, 
are shown using dash-dot-dot lines.
\label{F:FIG3b}
}
\end{figure}

\begin{figure}
\figurenum{3c}
\centerline{
\plotone{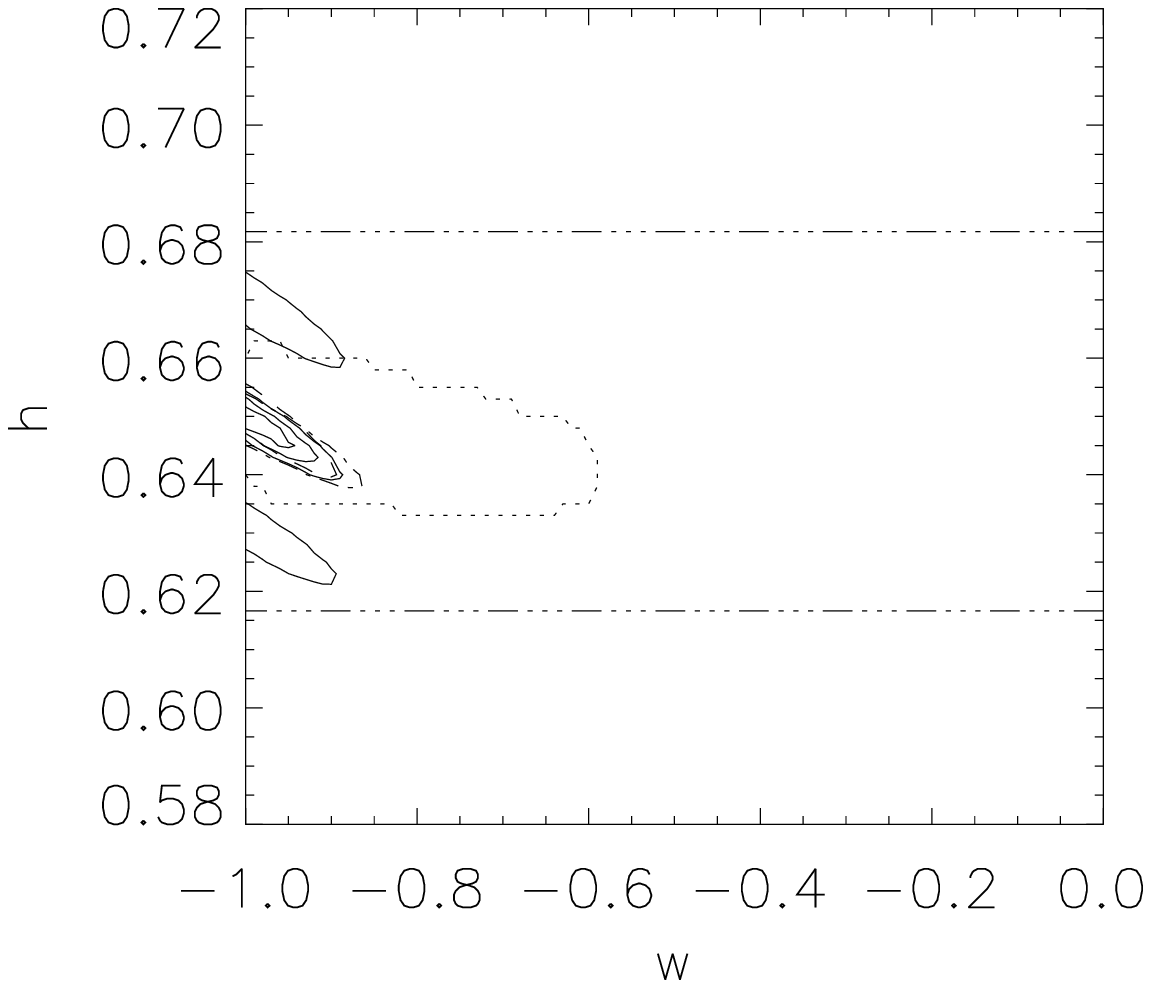}
}
\caption{1, 2 and 3$\sigma$ constraints on cosmological parameters from
simulated angular diameter distance measurements. Contours and lines
have the same meaning as in Figure 2. The figures show results of 
simulations using 500 clusters assuming an error of 4$\%$ in the
angular diameter distance (concentric ellipses, solid lines). 
3$\sigma$ constraints assuming an additional $\pm 3\%$ systematic 
error are shown using solid ellipses above and below the random
ellipses (panels b and c). Panels a, b, and c show two dimensional
slices of the three dimensional parameter space defined by
($\Omega_m$, $w$, and $h$) and passing through the best-fit model
as in Figure 2. Constraints from $\Omega_m h^2$ = constant,
assuming a 10$\%$ error in its determination from CMB experiments, 
are shown using dash-dot-dot lines.
\label{F:FIG3c}
}
\end{figure}

In Figures~\ref{F:FIG3}, b, and c we show similar constraints on
$\Omega_m$, $w$, and $h$ from the SZ/X-ray method using the same fiducial 
CDM model, and 500 clusters with a random error of 4$\%$ in $D_A$, 
as before (1, 2, and 3$\sigma$ concentric ellipses, solid lines).
The constraints form a banana-shaped region elongated in
the $\Omega_m$ - $w$ plane (Figure~\ref{F:FIG3}). 
Systematic errors in the $D_A$ will not affect the results in the 
$\Omega_m - w$ plane, but simply shift the ellipses from random errors 
up or down along the $h$ axis
(Figures~\ref{F:FIG3b} and c, 3$\sigma$ solid ellipses above and under
the 1, 2, and 3$\sigma$ concentric ellipses of random errors, solid lines).
Again, with the assumed 4$\%$ random and $\pm 3\%$ systematic errors,
the $h$ determination is going to be limited by systematic errors.
We also carried out simulations adding 
a systematic error with a gradient in redshift growing from 0$\%$ at $z=0$
to $\pm 3\%$ at $z=1$ to the assumed 4$\%$ random error
(Figures~\ref{F:FIG3b} and c, short dashed and dash dotted 3$\sigma$ lines).
Again, we find that $\Omega_m$, and $w$ are strongly affected, 
their determination is limited by the assumed systematic errors. 
As before, the Hubble constant determination 
is not affected by this type of systematic errors.
Also, an indication of systematic errors in $D_A$ from evolution might 
be found using a $w$ - $h$ plot (Figure~\ref{F:FIG3c}).

From Figure~\ref{F:FIG2c}, which shows constraints in the
$\Omega_\Lambda$ - $h$ plane, we can conclude that with the assumed
accuracy, the cosmological constant is not well constrained.
From Figure~\ref{F:FIG3} we can see that $w$ is also poorly constrained 
by the SZ/X method. Clearly, whichever set of parameters
is to be estimated, other cosmological measurements are needed to
constrain these parameters further.

One of the most promising other experiments to measure cosmological
parameters is based on CMB fluctuations. We estimated how well we can
determine cosmological parameters if we add constraints from CMB
experiment, which we would expect to be particularly useful since 
the strongest dependency in this experiment is on the space curvature
(the total average density in the Universe). As an illustration of the
power of combining these techniques, we used the position of the 
first Doppler peak in the angular power spectrum of CMB fluctuations
\citep{Hanet00,Mauet00,Milet99}, and the combination $\Omega_m h^2$ = constant
as observables from primordial fluctuation studies
(see for example Zaldarriaga, Spergel \& Seljak 1997).

The position of the first Doppler peak can be expressed as
\begin{equation} \label{E:}
  \ell_{peak} = k_{peak} r(z_*)
,
\end{equation}
where $k_{peak}$ and $r(z_*)$ are the first peak in $k$ space and the
effective distance to the last scattering surface at $z_*$
\begin{equation} \label{E:}
r(z_*) = { 1 \over \sqrt{ K} } \; {\cal S} \;
                   \Bigl[\sqrt{ K} \bigl( \eta(0) - \eta(z_*) \bigr) \Bigr]
,
\end{equation}
where $K$ is the curvature, $\cal S$ is the $\sin$, $\sinh$, and identity function for
models with negative, positive and flat space-time, and $\eta$ is the conformal time.
In $k$-space in a CDM model, 
\begin{equation} \label{E:}
  k_{peak} \approx c_1 + c_2 w_m + c_3 w_m^2 + c_4 w_b 
                   + c_5 w_b^2 + c_6  w_m w_b
,
\end{equation}
where $w_m = \Omega_m h^2$ and $w_b = \Omega_b h^2$,
and the coefficients are: $c_1 = 0.0112 $, $c_2 = 0.0441$, 
$c_3 = -0.043$, $c_4 = -0.0496$, 
$c_5 = 2.65$, and $c_6 = 0.162$ \citep{Whit98}.
We used \cite{HuSug96}'s approximation for the redshift of the last
scattering surface
\begin{equation} \label{E:}
  z_* = 1048 [ 1 + 0.0012 w_b^{-0.738}] [ 1 + g_1^{g_2}]
  ,
\end{equation}
where
\begin{equation} \label{E:}
 g_1 = { 0.0783 w_m^{-0.238} \over  1 + 39.5 w_m^{0.763} }
 ,
\end{equation}
and 
\begin{equation} \label{E:}
 g_2 = { 0.560 \over 1 + 21.1 w_b^{1.81} }
 ,
\end{equation}

In Figures~\ref{F:FIG2}, b and c we over-plot the 3$\sigma$ error
region based on the position of the first Doppler peak in the CMB
power spectrum (long dashed lines).
This region is roughly aligned along the line
$\Omega_m + \Omega_\Lambda$ = constant because of the strong
dependence of the position of the first Doppler peak on the total
space curvature. For demonstration purposes, we choose 
$\ell_{peak} = 245 \; \pm 10$, where we assume a $3\sigma$ error range of 
$\Delta \ell = 10$. As expected, constraints on cosmological
parameters from SZ/X distance-redshift relation and the position of
the first peak in the CMB fluctuations are highly complementary. For
our assumed set of clusters, uniformly sampled in $z < 1$,
these two constraints are nearly orthogonal to each other in 
the $\Omega_m$ - $\Omega_\Lambda$ plane (Figure~\ref{F:FIG2}).
The constraints are also complementary in the
$h - \Omega_m$ and $h - \Omega_\Lambda$ planes (Figures~\ref{F:FIG2b} 
and c). Geometrically, we have narrow banana-shaped constraints
from the SZ/X-ray method elongated in the $\Omega_m$ -
$\Omega_\Lambda$ plane and narrow sheet-like constraints $h = \rm
constant$ from the position of the first peak of the CMB fluctuation
spectrum. 
The intercept of these constraints gives us stringent constraints on
the following cosmological parameters: $\Omega_m$, $\Omega_\Lambda$, 
and $h$ can be determined to $\pm 0.08$, $\pm 0.1$, and $\pm 0.015$
($3\sigma$, assuming 4$\%$ random error in $D_A$).
Note, that while the effect on $\Omega_m$ and $\Omega_\Lambda$ 
from reasonable redshift independent systematic errors are negligible, 
$h$ would be dominated by systematic errors: 
a $\pm 3\%$ systematic error would result a $\pm 0.02$ change in $h$.
A systematic error which grows from 0$\%$ to $\pm 3\%$ at $z=1$ would 
cause an additional $\pm 0.05$, $\pm 0.05$, and $\pm 0.005$ error in 
$\Omega_m$ and $\Omega_\Lambda$, and $h$.
Constraints from the SZ/X-ray method are also orthogonal
to those from cluster evolution (compare constraints on the $\Omega_m$
- $\Omega_\Lambda$ plane, our Figure~\ref{F:FIG2} and Figure 1 of
Holder et al. 2001). 
In Figures~\ref{F:FIG2}, b and c
we also over-plot constraints from $\Omega_m h^2$ = constant
assuming a 10$\%$ error in its determination from CMB experiments 
(dash dot dot dotted lines), as suggested by studies based on the 
characteristics of the MAP experiment \citep{Zalet97}.
These constraints seem to be less useful in this parameter space 
than those from the position of the first Doppler peak.

Constraints from the location of the first peak in the CMB
fluctuations are not so useful when considering the alternative set
of parameters $w$, $\Omega_m$, $h$. The $\ell \approx 210-240$
constraint leads to surfaces which lie almost parallel to the 
ellipsoids derived from the SZ/X-ray method
(compare our Figure~\ref{F:FIG3} and Figure 4 of White 1998). 
In Figures~\ref{F:FIG3}, b, and c we over-plot constraints from 
$\Omega_m h^2$ = constant again assuming a 10$\%$ error in its 
determination (dash dot dot dotted lines).
Combining constraints from the SZ/X-ray method and those from
$\Omega_m h^2$ = constant (based on CMB fluctuation analysis), 
we obtain stringent constraints on $w$ and $h$
($\le 0.2$ and $\pm 0.015$, $3\sigma$).
Note, that the effect on $w$ from redshift independent systematic 
errors is negligible, and that, as before, $h$ is dominated by 
systematic errors:
a $\pm 3\%$ systematic error would result a $\pm 0.02$ change in $h$.
A systematic error which grows from 0$\%$ to $\pm 3\%$ at $z=1$ would 
cause an additional $\pm 0.1$ in $w$.
From Figure~\ref{F:FIG3b} ($\Omega_m - h$ plane), we see that a 
gradient in the systematic error in redshift might be recognized 
using the $\Omega_m h^2$ = constant constraint.
Also, constraints from the shape of the
power spectrum of CMB fluctuations (as will be achieved by MAP and
Planck) are nearly orthogonal to those from the SZ/X-ray method
(compare our Figure~\ref{F:FIG3} and Figure 13 of Huterer and Turner
2000). Comparing our Figure~\ref{F:FIG3} to Figures 8 and 9 of
\cite{Haimet00}, we can see that constraints on the $\Omega_m$ - $w$
plane from distance function and cluster evolution are complementary.
Also, constraints on $w$ from the SZ/X-ray method combined with cluster
abundance would put stringent constraints on $w$.

Since the banana-shaped constraints from the SZ/X method in the
parameter spaces defined by ($\Omega_m$, $\Omega_\Lambda$, $h$) or
($\Omega_m$, $w$, $h$) are nearly orthogonal to constraints from cluster 
evolution, a strategy for determining cosmological constants based {\it
only} on clusters is likely to be successful. It seems to be possible
to choose the parameters of the set of clusters used in this work such
that constraints from the SZ/X ray method and cluster evolution can be
made orthogonal, and thus optimized for separating cosmological
parameters. Clusters can provide a powerful, independent test for
cosmological parameters.

Note that our constraints in the $\Omega_m$ - $w$ plane are curved towards 
the $w$ axis (Figure~\ref{F:FIG3}, solid and dotted lines),
while Huterer \& Turner (2000)'s constraints from SNe Ia (which is
basically a constraints from distance - redshift function, cf. their
Figure 23) show no sign of curvature. This is due to the fact that
they used the Fisher matrix formalism, and we evaluated the $\chi^2$
statistic directly. Our results show that the likelihood function is
strongly non-Gaussian on the $\Omega_m$ - $w$ plane.

As a second realization, we carried out simulations with seventy clusters 
(35 high redshift clusters, $z > 0.5$), a number likely to be observed in 
the next few years, and estimated how well we can constrain cosmological 
parameters. We assumed the same fiducial cosmological model as before: 
$\Omega_m=0.3$, $\Omega_\Lambda=0.7$, and $h$ = 0.65. 
For this simulation, however, we assumed a random error of 7$\%$ in the 
angular diameter distance, which might be achievable in the near future
(we assume lensing measurements will be used to eliminate errors from
cluster shape).

In Figures~\ref{F:FIG4}, b, and c, we show the resulting 1, 2 and
3$\sigma$ constraints on the parameter space defined by ($\Omega_m$,
$\Omega_\Lambda$, $h$), corresponding to $\Delta \chi^2$ of 3.53, 8.02 
and 14.2 appropriate for three fitted parameters (solid lines). 
Long dashed lines show the constraints 
from the position of the first Doppler peak. As before, we assumed
$\ell_{peak} = 245 \; \pm 10$ ($3\sigma$ range). 
We also over-plot constraints from $\Omega_m h^2$ = constant
assuming a 10$\%$ error in its determination from CMB experiments 
(dash dot dot dotted lines).
From these Figures we conclude that by using as few as 35
high-redshift (and 35 low-redshift) clusters, with a random error of
7$\%$ in $D_A$ we can constrain $\Omega_m$, $\Omega_\Lambda$ and $h$
within $\pm 0.2$, $\pm 0.2$, and $\pm 0.04$ ($3\sigma$ errors).
Note, that the effect of redshift independent systematic errors in $D_A$ 
on $\Omega_m$ and $\Omega_\Lambda$, occurring in practice, is negligible,
but a $\pm 5\%$ systematic error would result an additional $\pm 0.035$
error in $h$. 
Assuming that we know that our redshift independent systematic error is
less than 15$\%$ at the 3$\sigma$ level, with a 7$\%$ random error
combined in quadrature, we would be able to determine $h$ with an error
of $\pm 0.11$.
A systematic error which grows from 0$\%$ to $\pm 5\%$ at $z=1$ would 
cause an additional $\pm 0.1$, $\pm 0.1$ and $\pm 0.01$ error in
$\Omega_m$, $\Omega_\Lambda$, and $h$
(Figure 4, short dashed and dash dotted lines).
This means that as few as seventy clusters, with the errors likely 
to be achieved in the next few years, would be sufficient to exclude 
models with zero cosmological constant with high significance. 
These results would be independent of the SNe Ia data, thus they would 
provide a robust check to the supernova results.

In Figures~\ref{F:FIG5}, b, and c, we show the 1, 2 and 3$\sigma$ constraints
on the parameter space defined by ($\Omega_m$, $w$, $h$), 
with the $\Omega_\Lambda = 1 - \Omega_m$ constraint (solid lines). 
We over-plot constraints from $\Omega_m h^2$ = constant assuming a 
10$\%$ error as before (dash dot dot dotted lines).
From Figure~\ref{F:FIG5} we can see that 
we will be able to determine the equation of state parameter, $w$, 
with accuracy of 0.45 (3$\sigma$), even with reasonable systematic errors
in $D_A$, with additional information from CMB experiments, or 
SZ effect number counts.
A systematic error which grows to $\pm 5\%$ at z = 1 would cause an additional
error of 0.2 in $w$ (Figure 5, short dashed and dash dotted lines).

% % % % % % % % % % % % % % % % % % % % % % % % % % % % % % % % % % % % % % % % % % % % % % % % %
\section{Conclusion}
\label{S:Conclusion}

We have estimated the accuracy achievable in the determination of
cosmological parameters using the SZ/X-ray method of distance determination.
This method uses a sample of clusters to map the distance-redshift relation, 
which is a sensitive probe of cosmological parameters.
The advantages of this well-known method are: 
(1) unlike other distance determination methods,
 it depends only on the geometry of the Universe, and its average densities;
(2) it is a physical method, based on relatively simple gravitational 
 virialization of clusters (as opposed to complicated physics and chemistry
 involved in galaxy formation and supernova explosions); and
(3) a large number of clusters is available for observation,
 and thus systematic effects can be reduced by using many clusters, 
 or selecting clusters appropriately to reduce systematics.
The necessary data should be available within the next few years.

The SZ/X method, as other cosmological tests, 
can constrain well only some combination of cosmological parameters.
We have shown that constraints on ($\Omega_m$, $\Omega_\Lambda$, $h$)
from the $D_A(z)$ function measured in $z = 0 - 1$ are nearly
orthogonal to constraints from the first peak of the CMB fluctuations,
and also to constraints from cluster evolution. 
Constraints on ($\Omega_m$, $w$, $h$) from the $D_A(z)$ function are
complementary to those from cluster evolution (compare our
Figure 2 to Figures~8 and 9 from Haiman et al. 2000).
In general, $D_A$ provides constraints on cosmological parameters similar 
to those from the SNe Ia method. Constraints from cluster evolution 
(the source counts $N_{SZ}(z)$ and $N_{X}(z)$) are similar to
constraints from the full CMB fluctuation spectrum as will be measured
by MAP and Planck. 

This result suggests that cosmological tests using {\it
only} clusters of galaxies can be an important
independent check on the values of cosmological parameters measured by
other techniques. Cluster-based methods have systematic errors which
are unrelated to those of other methods of measuring 
cosmological parameters (e.g., from CMB fluctuations, and
SNe~Ia). Furthermore, when the set of cosmological parameters is
further extended, with additional components of density or structure
parameters, joint analyses using multiple cosmological tests will be
essential to remove the parameter degeneracy exhibited by any one test.

We demonstrated the effect of systematic errors on the determination
of cosmological parameters. If random errors can be kept at a few
percent level, systematic errors of similar magnitude will dominate
the error in the Hubble constant. $\Omega_m$, $\Omega_\Lambda$, and $w$
are not affected by redshift-independent systematic errors. We also
showed that a systematic error with a linear gradient in redshift
will not affect Hubble constant determinations, but causes systematic
shifts in the estimates of $\Omega_m$, $\Omega_\Lambda$, and $w$.

\begin{figure}
\figurenum{4a}
\centerline{
\plotone{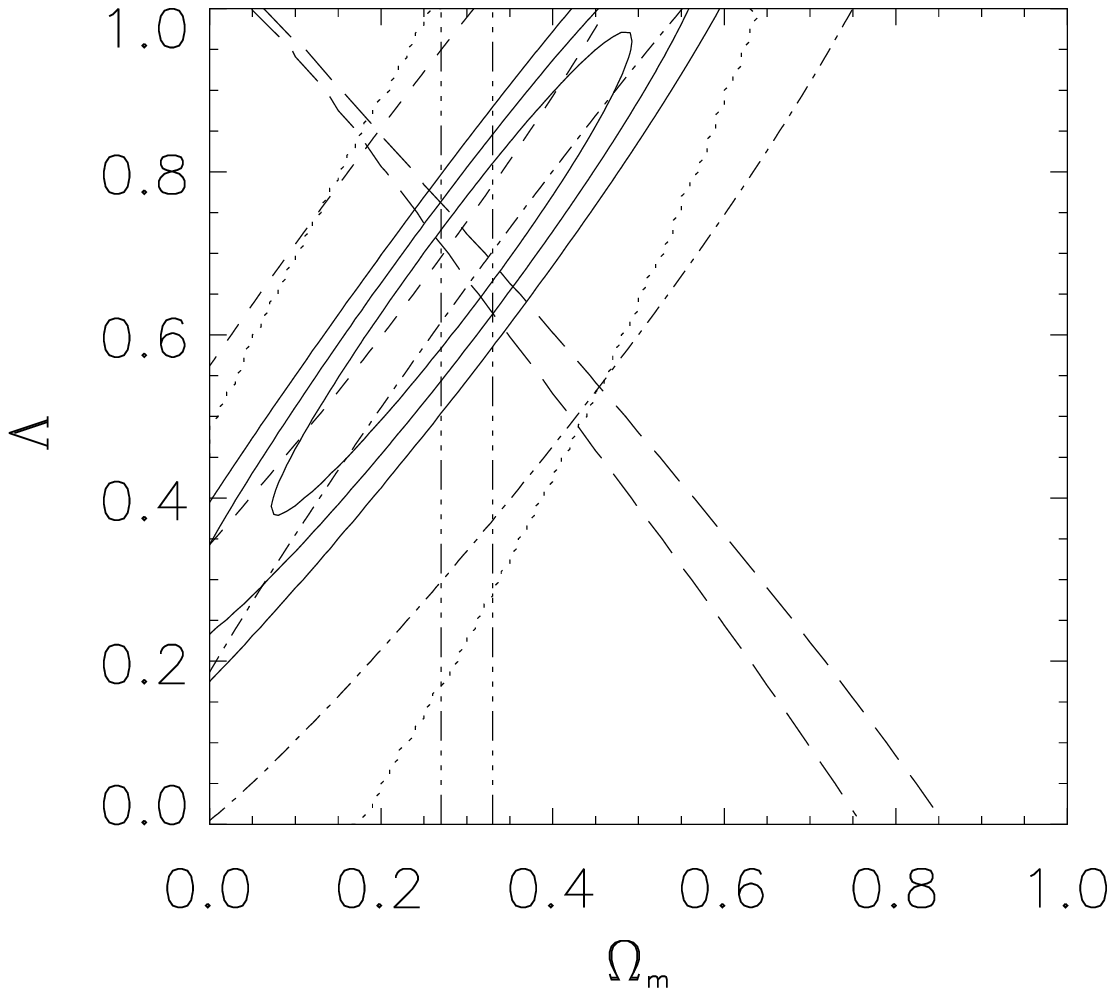}
}
\caption{Same as Figure 2, but with only 70 clusters (with 35 at $z > 0.5$)
assuming a 7\% random error in the angular diameter distances. We
omit the redshift-independent systematic error ellipses, since they
simply correspond to the same shifts on the $h$ axis seen in Fig.~2,
but retain the redshift-dependent systematic error ellipses
corresponding to a 5\% gradient in the error in $D_A$.
\label{F:FIG4}
}
\end{figure}

We showed, that in the near future, even with only 35 high redshift
(and 35 low redshift) clusters, the SZ/X-ray method, combined with
the position of the first peak in the power spectrum of CMB
fluctuations, will provide enough accuracy to exclude $\Omega_\Lambda
= 0$ models with high confidence even with constant and redshift 
dependent systematic errors (see Figure~\ref{F:FIG4}). 
Also, the SZ/X-ray method, combined with the $\Omega_m h^2$ = constant 
constraint, or cluster abundance, will allow us to determine the equation 
of state parameter, $w$, to within 0.45
(3$\sigma$; Figure~\ref{F:FIG5}), even with usual (redshift-independent) 
systematic errors.

This idealized discussion of the SZ/X-ray method and its errors in the
determination of cosmological parameters could be improved by using the
detailed characteristics of specific instruments and observing
strategies. Departures from the assumptions made here could cause
increases or decreases in the level of error in the derived
parameters, with the largest changes likely for different redshift
samplings.

We conclude, that the determination of the angular diameter distance -
redshift function using the SZ effect and X-ray thermal bremsstrahlung
emission from clusters of galaxies can be used with confidence to
constrain cosmological parameters. In general, clusters of galaxies
{\it alone} can be used to constrain cosmological parameters
independently from other methods. Clusters lead to 
parameter limits which are competitive with using other techniques.

%\acknowledgement

Most of this work was done while SMM held a National Research Council 
Research Associateship at NASA Goddard Space Flight Center. 
We thank David Spergel and the anonymous referee for 
useful comments and suggestions.

% % % % % % % % % % % % % % % % % % % % % % % % % % % % % % % % % % % % % % % % % % % % % % % % %
% 
%                                      B I B L I O G R A P H Y 
% 
% % % % % % % % % % % % % % % % % % % % % % % % % % % % % % % % % % % % % % % % % % % % % % % % %

% % % % % % % % % % % % % % % % % % % % % % % % % % % % % % % % % % % % % % % % % % % % % % % % %
% 
%                 F I G U R E S            F I G U R E S                 F I G U R E S          
% 
% % % % % % % % % % % % % % % % % % % % % % % % % % % % % % % % % % % % % % % % % % % % % % % % %

\clearpage

%  FIGURE 1

\clearpage

% % % % % % % % % % % % % % % % % % % % % % % % % % % % % % % % % % % % % % % % % % % % % % % % %
% Omega_M -  LAMBDA - H   500
% % % % % % % % % % % % % % % % % % % % % % % % % % % % % % % % % % % % % % % % % % % % % % % % %

%  FIGURE 2abc

\clearpage

%  FIGURE 2abc

\clearpage

%  FIGURE 2abc

\clearpage

% % % % % % % % % % % % % % % % % % % % % % % % % % % % % % % % % % % % % % % % % % % % % % % % %
% Omega_M -  W - H   500
% % % % % % % % % % % % % % % % % % % % % % % % % % % % % % % % % % % % % % % % % % % % % % % % %

%  FIGURE 3abc

\clearpage

%  FIGURE 3abc

\clearpage

%  FIGURE 3abc

\clearpage

% % % % % % % % % % % % % % % % % % % % % % % % % % % % % % % % % % % % % % % % % % % % % % % % %
% Omega_M -  LAMBDA - H   70
% % % % % % % % % % % % % % % % % % % % % % % % % % % % % % % % % % % % % % % % % % % % % % % % %

%  FIGURE 4abc

\clearpage

%  FIGURE 4abc
\begin{figure}
\figurenum{4b}
\centerline{
\plotone{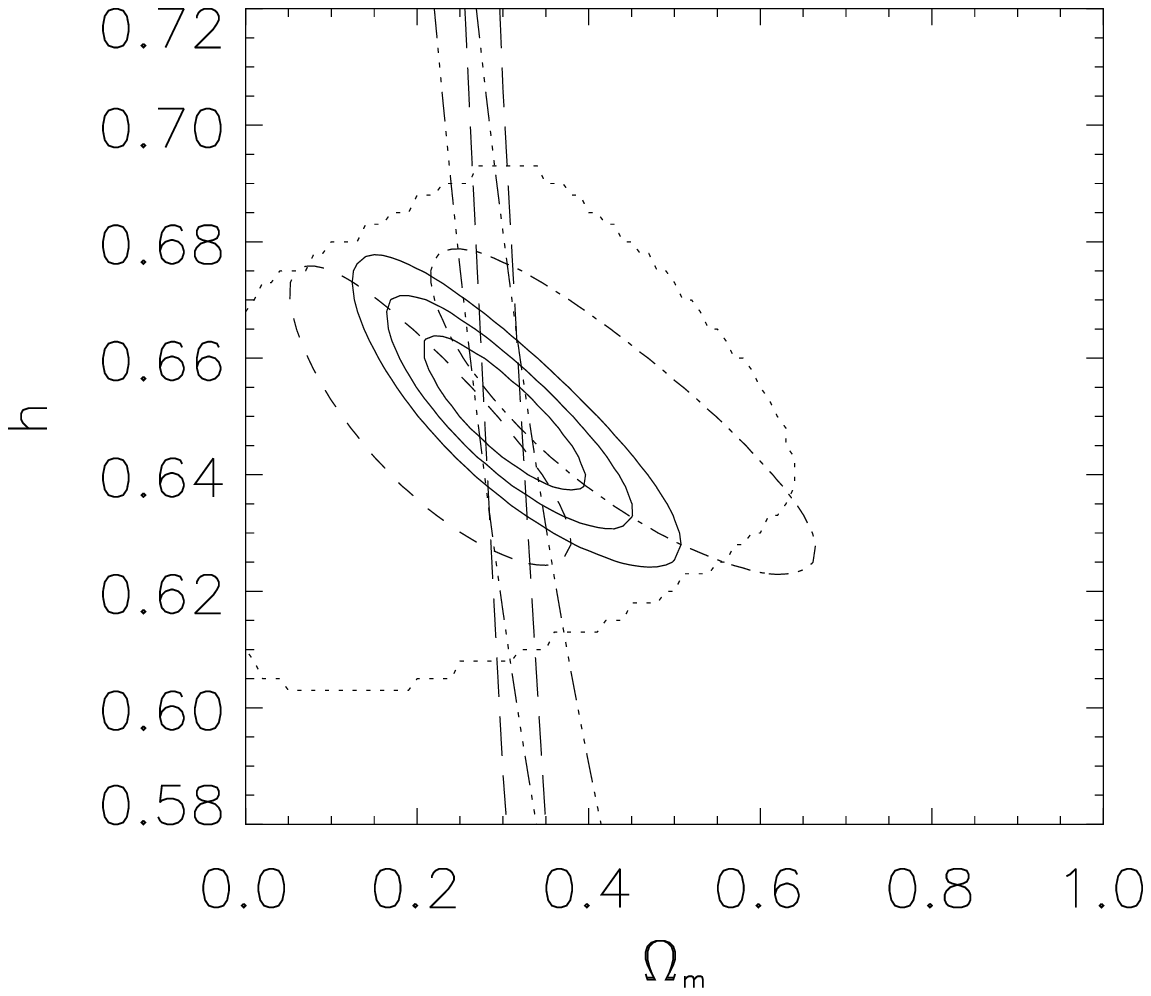}
}
\caption{Same as Figure 2, but with only 70 clusters (with 35 at $z > 0.5$)
assuming a 7\% random error in the angular diameter distances. We
omit the redshift-independent systematic error ellipses, since they
simply correspond to the same shifts on the $h$ axis seen in Fig.~2,
but retain the redshift-dependent systematic error ellipses
corresponding to a 5\% gradient in the error in $D_A$.
\label{F:FIG4b}
}
\end{figure}

\clearpage

%  FIGURE 4abc
\begin{figure}
\figurenum{4c}
\centerline{
\plotone{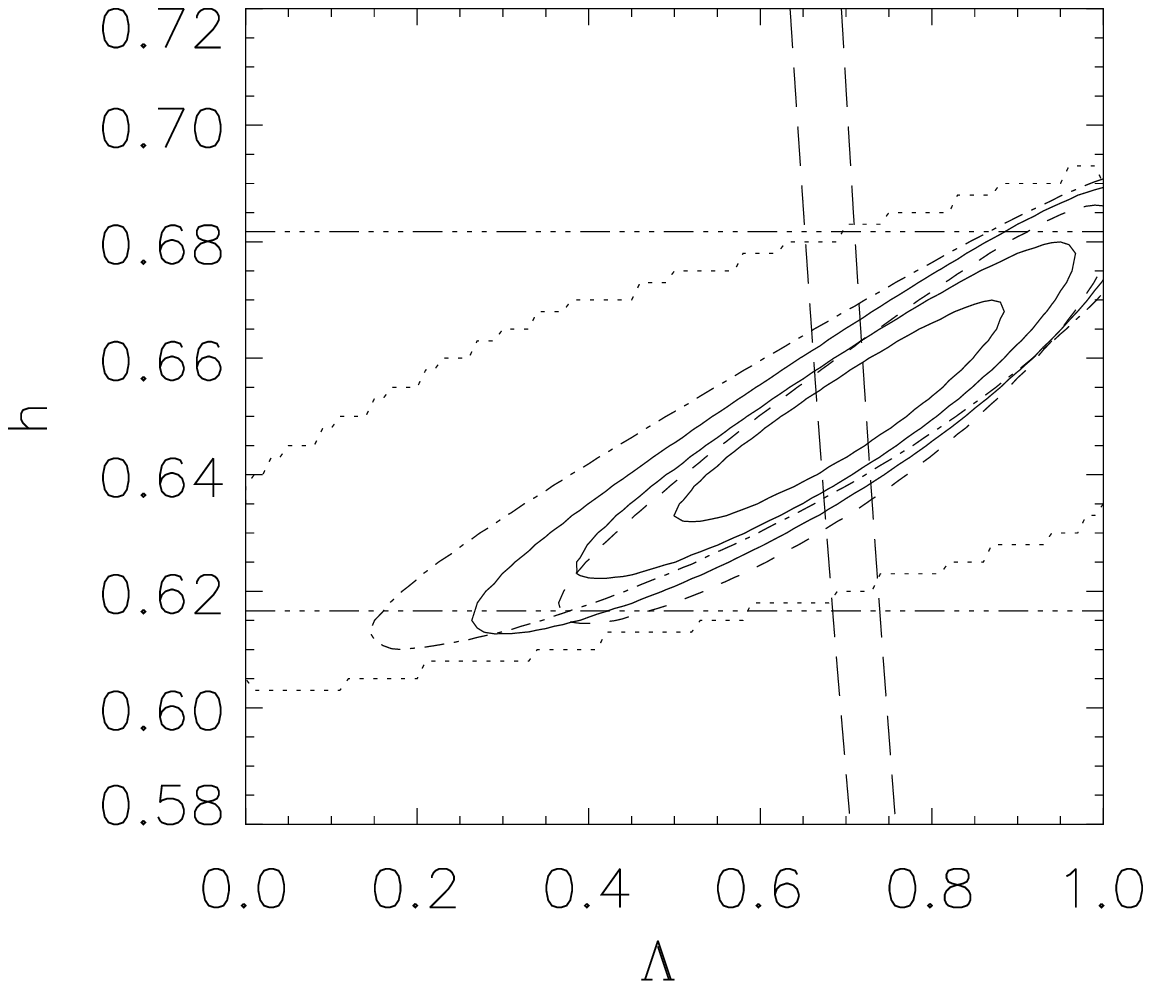}
}
\caption{Same as Figure 2, but with only 70 clusters (with 35 at $z > 0.5$)
assuming a 7\% random error in the angular diameter distances. We
omit the redshift-independent systematic error ellipses, since they
simply correspond to the same shifts on the $h$ axis seen in Fig.~2,
but retain the redshift-dependent systematic error ellipses
corresponding to a 5\% gradient in the error in $D_A$.
\label{F:FIG4c}
}
\end{figure}

\clearpage

% % % % % % % % % % % % % % % % % % % % % % % % % % % % % % % % % % % % % % % % % % % % % % % % %
% Omega_M -  W - H   70
% % % % % % % % % % % % % % % % % % % % % % % % % % % % % % % % % % % % % % % % % % % % % % % % %

%  FIGURE 5abc
\begin{figure}
\figurenum{5a}
\centerline{
\plotone{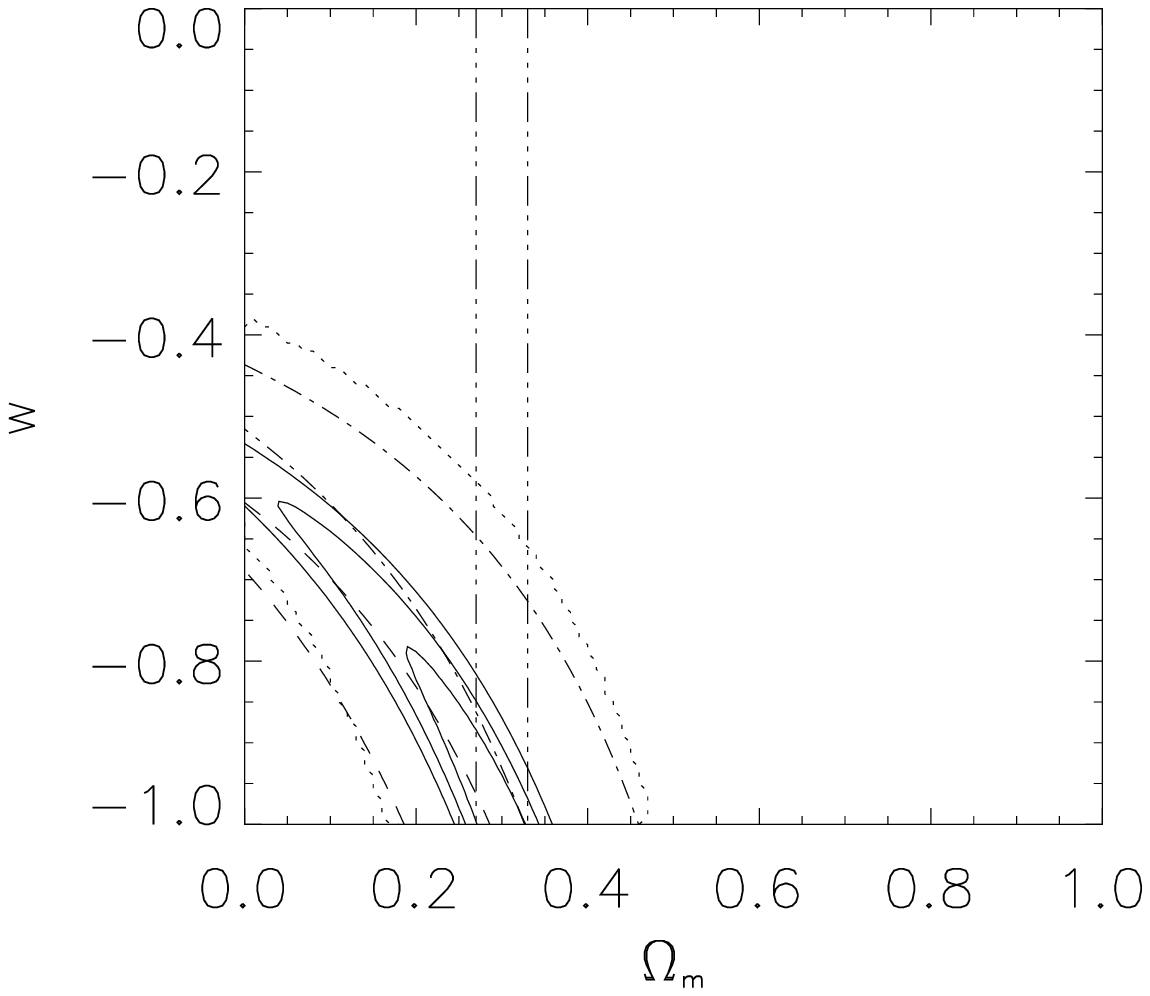}
}
\caption{Same as Figure 3, but with only 70 clusters (with 35 at $z > 0.5$)
assuming a 7\% random error in the angular diameter distances. We
omit the redshift-independent systematic error ellipses, since they
simply correspond to the same shifts on the $h$ axis seen in Fig.~2,
but retain the redshift-dependent systematic error ellipses
corresponding to a 5\% gradient in the error in $D_A$.
\label{F:FIG5}
}
\end{figure}%Diego, J. M., et al. 2001, astro-ph/0104217

\clearpage

%  FIGURE 5abc
\begin{figure}
\figurenum{5b}
\centerline{
\plotone{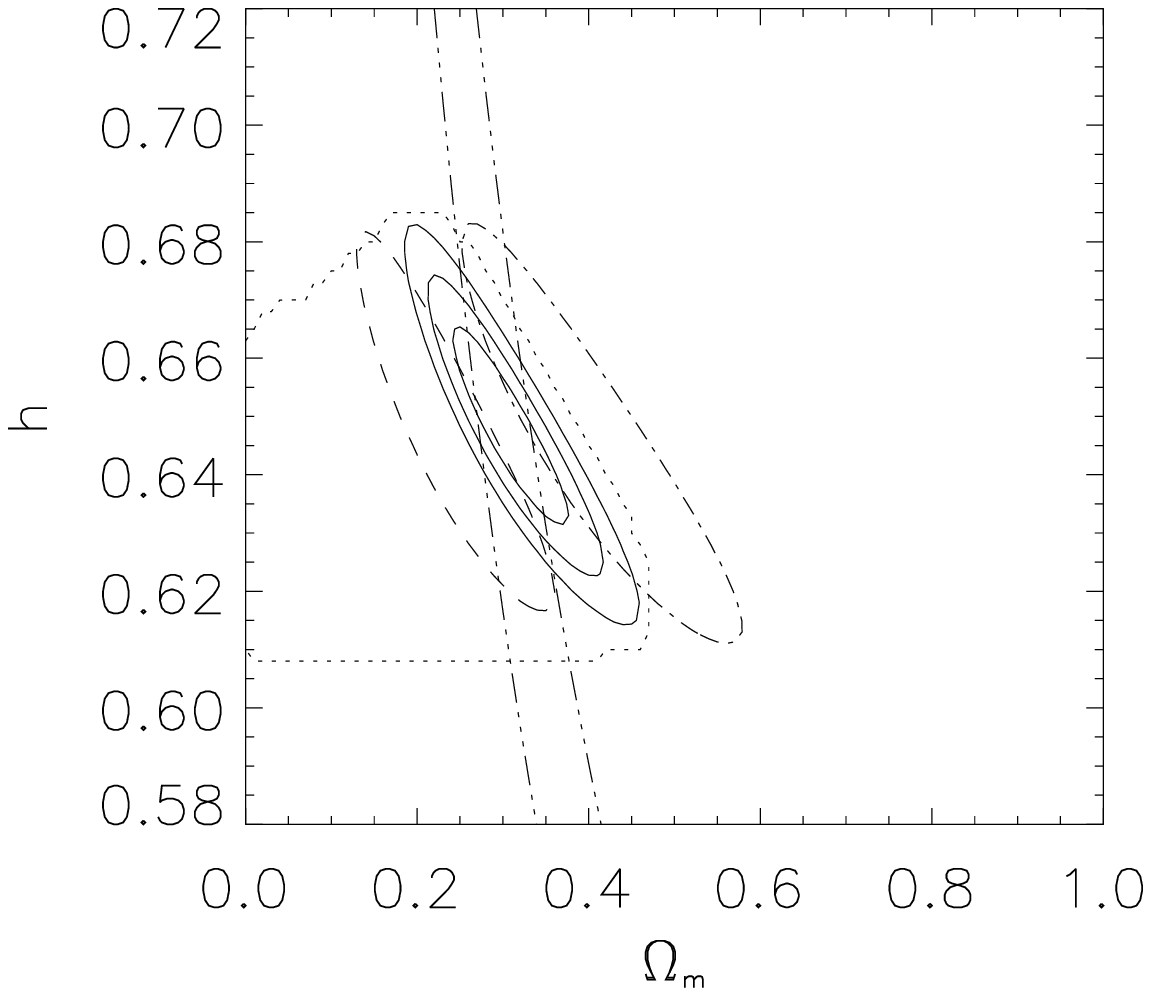}
}
\caption{Same as Figure 3, but with only 70 clusters (with 35 at $z > 0.5$)
assuming a 7\% random error in the angular diameter distances. We
omit the redshift-independent systematic error ellipses, since they
simply correspond to the same shifts on the $h$ axis seen in Fig.~2,
but retain the redshift-dependent systematic error ellipses
corresponding to a 5\% gradient in the error in $D_A$.
\label{F:FIG5b}
}
\end{figure}

\clearpage

%  FIGURE 5abc
\begin{figure}
\figurenum{5c}
\centerline{
\plotone{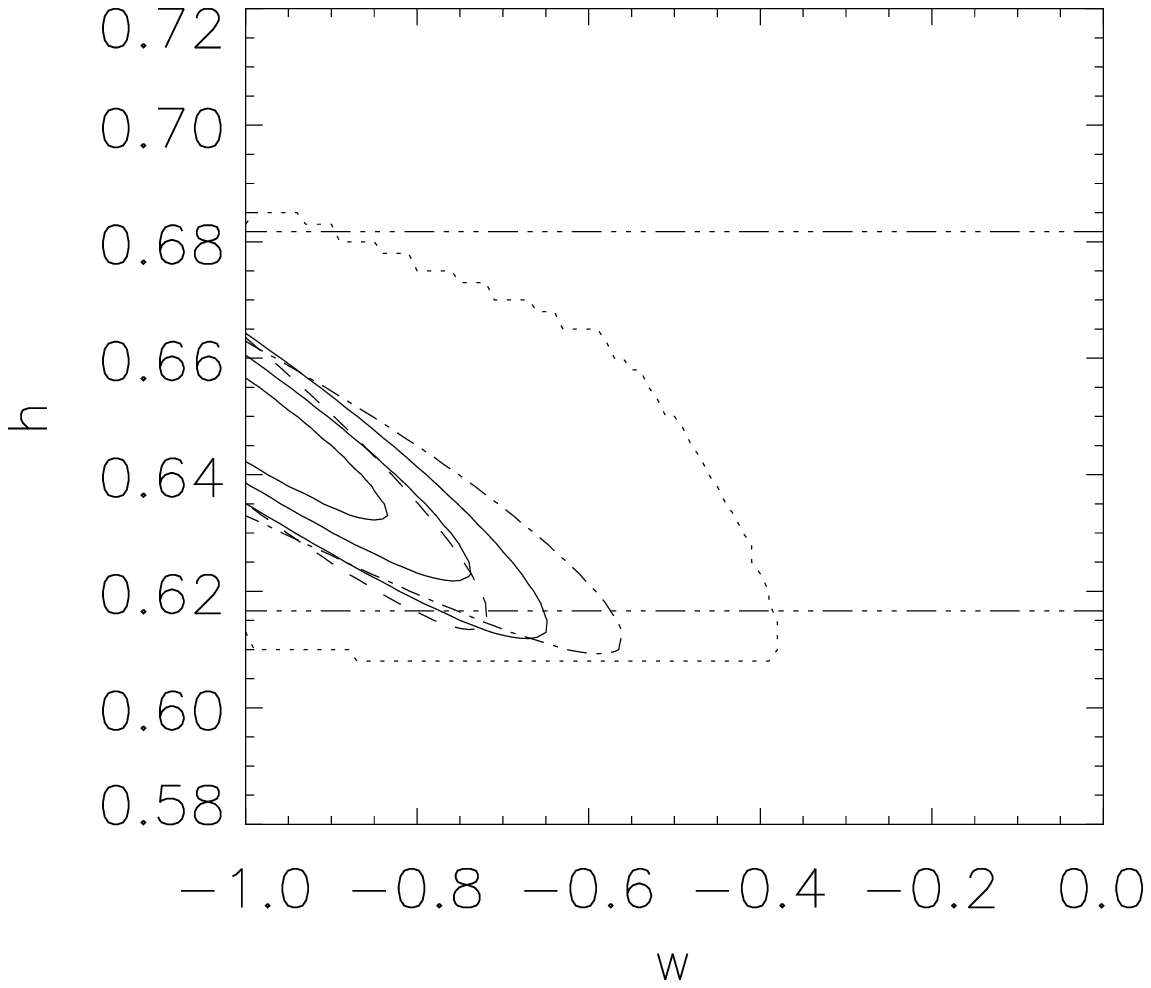}
}
\caption{Same as Figure 3, but with only 70 clusters (with 35 at $z > 0.5$)
assuming a 7\% random error in the angular diameter distances. We
omit the redshift-independent systematic error ellipses, since they
simply correspond to the same shifts on the $h$ axis seen in Fig.~2,
but retain the redshift-dependent systematic error ellipses
corresponding to a 5\% gradient in the error in $D_A$.
\label{F:FIG5c}
}
\end{figure}

% % % % % % % % % % % % % % % % % % % % % % % % % % % % % % % % % % % % % % % % % % % % % % % % %
% 
%                                    T H E          E N D         
% 
% % % % % % % % % % % % % % % % % % % % % % % % % % % % % % % % % % % % % % % % % % % % % % % % %

\begin{thebibliography}{}


\bibitem[Arnaud et al. (2001)]{Arnaet01}
Arnaud, M., Neumann, D. M., Aghanim, N., Gastaud, R., Majerowicz, S., 
\& Hughes, J. P., 2001, \aap, 365, L80

\bibitem[Allen \& Fabian (1998)]{AlleFabi98}
Allen, S. W., \& Fabian, A. C., 1998, \mnras, 297, L63

\bibitem[Balbi et al. (2000)]{Balbet00}
Balbi, A., et al. 2000, astro-ph/0005124 

\bibitem[Bahcall \& Soneira (1983)]{BahcSone83}
Bahcall, N. A., \& Soneira, R. M., 1983, \apj, 270, 20

\bibitem[Basilakos et al. (2000)]{BasPM00} 
Basilakos, S., Plionis, M., \& Maddox, S. J., 2000, astro-ph/0002459

\bibitem[Bartlett (2000)]{Bart00}
Bartlett, J. G., 2000, astro-ph/0001267

\bibitem[Bernardis et al. (2001)]{Bernet01}
de Bernardis, P. et al., 2001, astro-ph/0105296

\bibitem[Birkinshaw (1979)]{Birk79}
Birkinshaw, M., 1979, \mnras, 187, 847

\bibitem[Birkinshaw (1999)]{Birk99}
Birkinshaw, M., 1999, Physics Reports, 310, 97

\bibitem[Birkinshaw \& Hughes (1994)]{BirkHugh94}
Birkinshaw, M., \& Hughes, J. P., 1994, \apj, 420, 33

\bibitem[Birkinshaw et al. (1991)]{Birket991}
Birkinshaw, M., Hughes, J. P., \& Arnoud, K. A., 1991, \apj, 379, 466

\bibitem[Blain (1998)]{Blai98}
Blain, A. W., 1998, \mnras, 297, 502

\bibitem[Blasi (2000)]{Blas00}
Blasi, P., 2000, astro-ph/0008113

\bibitem[Bridle et al. (1999)]{Bridet99}
Bridle, S. L., et al., 1999, astro-ph/9903472

\bibitem[Browne et al. (2000)]{Brownet00}
Browne, I.W.A. et al. 2000. SPIE, 4015, 299

\bibitem[Carlstrom et al. (2001)]{Carlet01}
Carlstrom, J. E., et al., 2001, astro-ph/0103480

\bibitem[Cavaliere \& Fusco-Femiano (1976)]{CavaliereF76} 
Cavaliere, A., \& Fusco-Femiano, R., 1976, \aap, 49, 137

\bibitem[Cavaliere et al. (1979)]{Cavalet79}
Cavaliere, A., Danese, L., \& De Zotti G., 1979, \aap, 75, 322

\bibitem[Cen (1998)]{Cen_98}
Cen, R, 1998, \apjl, 498, L99

\bibitem[Challinor \& Lasenby (1998)]{ChalLase98}
Challinor, A., \& Lasenby, A., 1998, \apj, 499, 1

\bibitem[Colafrancesco (1999)]{Colaf99}
Colafrancesco, S., 1999, astro-ph/9907329

\bibitem[Colberg et al. (2000)]{Col00}
Colberg, J. M., et al., 2000, \mnras, 313, 229 

\bibitem[Cole et al. (1994)]{Coleet94}
Cole, S., Fisher, K. B., \& Weinberg, D. H., 1994, \mnras, 267, 785

\bibitem[David et al. (2001)]{Daviet01}
David, L. P., et al., 2001, astro-ph/0010224

\bibitem[De Grandi \& Molendi (1999)]{DeGrMole99}
De Grandi, S., \& Molendi, S., 1999, astro-ph/9911039

\bibitem[Diego et al. (2001)]{Dieget01}
Diego, J. M., Martinez-Gonzales, E., Sanz, J. L., Benitez, N., \& Silk, J., 
2001, astro-ph/0103512

\bibitem[Efstathiou \& Bond (1999)]{EfstBon99}
Efstathiou, G., \& Bond, J. R., 1999, MNRAS, 304, 75

\bibitem[Efstathiou et al. (1999)]{Efstet99}
Efstathiou, G., Bridle, S. L., Lasenby, A. N., Hobson, M. P., \& Ellis, R. S., 
1999, MNRAS, 303, L47

\bibitem[Evrard et. al. (1994)]{Evret94}
Evrard, A. E., Summers, F. J., \& Davis, M., 1994, \apj, 422, 11 

\bibitem[Fabian (1994)]{Fabi94}
Fabian, A. C., 1994, ARAA, 32, 277

\bibitem[Fusco-Femiano et al. (2001)]{Fusc01}
R.Fusco-Femiano, R., D.Dal Fiume, D., M.Orlandini, M., G.Brunetti, G., L.Feretti, L., 
\& G.Giovannini G., 2001, astro-ph/0105049

\bibitem[Gawiser \& Silk (1998)]{GewiSilk98}
Gawiser, E., \& Silk, J., 1998, Science, 280, 1405

\bibitem[Gomez et al. (2000)]{Gomeet00}
Gomez, P. L., Hughes, J. P., \& Birkinshaw, M., 2000, astro-ph/0004263

\bibitem[Grainge et al. (1999)]{Grainet99}
Grainge, K. et al., 1999, astro-ph/9904165 

\bibitem[Guerra, Daly \& Wan (2000)]{Gueret00}
Guerra, E. J., Daly, R. A., \& Wan, L. 2000, astro-ph/0006454

\bibitem[Gunn \& Thomas (1996)]{GunnThom96}
Gunn, K. F., \& Thomas, P. A., 1996, \mnras, 281, 1133

\bibitem[Haiman et al. (2000)]{Haimet00}
Haiman, Z., Mohr, J. J., \& Holder, G. P., 2000, astro-ph/0002336

\bibitem[Hanany et al. (2000)]{Hanet00}
Hanany, S., et al., 2000,, \apj, 545, L5

\bibitem[Holder et al. (2000)]{Holdet00}
Holder, G.P., Mohr, J.J., Carlstrom, J.E., Evrard, A.E., \& 
 Leitch, E.M., 2000, \apj, 544, 629

\bibitem[Holder et al. (2001)]{Holdet01}
Holder, G. P., Haiman, Z., \&, Mohr, J. J., 2001, astro-ph/0105396

\bibitem[Holzapfel et al. (1997)]{Holzet97}
Holzapfel, W. L., et al. 1997, \apj, 480, 449

\bibitem[Hu \& Sugiyama (1996)]{HuSug96}
Hu, W., \& Sugiyama, N., 1996, \apj, 471, 542

\bibitem[Hu et al. (2000)]{Hu__et00}
Hu, W., Fukugita, M., Zaldarriaga, M., \& Tegmark, M., 2000, astro-ph/0006436

\bibitem[Hughes \& Birkinshaw (1998)]{HughBirk98}
Hughes, J. P., \& Birkinshaw, M., 1998, \apj, 501, 1

\bibitem[Huterer \& Turner (2000)]{HuteTurn00}
Huterer, D., \& Turner, M. S., 2000, astro-ph/0012510

\bibitem[Inagaki et al. (1995)]{Inaget95}
Inagaki, Y., Suginohara, T., \& Suto, Y., 1995, \pasj, 47, 411

\bibitem[Irwin \& Bregman (2000)]{IrwiBreg00}
Irwin, J. A., \& Bregman, J. N., 2000, astro-ph/0003123

\bibitem[Jaffe et al. (2000)]{Jaffet00}
Jaffe, A. H., et al., 2000, astro-ph/0007333

\bibitem[Jones et al. (2001)]{Joneet00}
Jones, M. E., et al. 2001, astro-ph/0103046

\bibitem[Kneissl et al. (2001)]{Kneiet01}
Kneissl, R., et al., 2001, astro-ph/0103042

\bibitem[Komatsu et al. (2001)]{Komet01}
Komatsu, E., et al., 2001, PASJ, 53, 57

\bibitem[Lasenby et al. (1999)]{Laseet99}
Lasenby,A. N., Bridle, S. L., \& Hobson, M. P., 1999, astro-ph/9901303

\bibitem[Lauer \& Postman (1994)]{LauePost94}
Lauer, T. R., \& Postman, M., 1994, \apj, 425, 418

\bibitem[Lee et al. (2001)]{Lee_et01}
Lee, A. T., et al., 2001, astro-ph/0104459

\bibitem[Lineweaver (1998)]{Line98}
Lineweaver, C., 1998, \apj, in press, astro-ph/9805326

\bibitem[Loeb \& Refregier (1997)]{LoebRefr97}
Loeb, A., \& Refregier, A., 1997, \apjl, 476, L59

\bibitem[Majumdar \& Subrahmanyan (2000)]{MajuSubr00}
Majumdar, S., \& Subrahmanyan, R., 2000, \apj, 542, 597

\bibitem[Maloney \& Bland-Hawthorn (2001)]{MaloBlan01}
Maloney, P. R., \& Bland-HawthoSstudiesrn, J., 2001, astro-ph/0104422

\bibitem[Mason (1999)]{Maso99}
Mason, B., 1999, PhD Thesis, Universtity of Pennsylvania

\bibitem[Mason et al. (2001)]{Masoet01}
Mason, B. S., Myers, S. T., \& Readhead, A. C. S., 2001, astro-ph/0101169 

\bibitem[Metcalf \& Silk (1998)]{MetcSilk98}
Metcalf, R. B., \& Silk, J., 1998, \apjl, 492, L1

\bibitem[Mauskopf et al. (2000)]{Mauet00}
Mauskopf, P. D.,  et al., 2000, \apj, 536, L59

\bibitem[Miller et al. (1999)]{Milet99}
Miller, A. D. et al., 1999, \apj, 524, 1

\bibitem[Mohr et al. (1995)]{Mohret95}
Mohr, J. J., Evrard, A. E., Fabricant, D. G., \& Geller, M. J., 1995, \apj, 447, 8

\bibitem[Mohr et al. (1999)]{Mohret99}
Mohr, J. J., Mathiesen, B., \& Evrard, A. E., 1999, \apj, 517, 627

\bibitem[Molnar (2000)]{SMM00}
Molnar, S. M., 2000, ASP Conference Series, eds.: Giacconi, R., Stella, L, \& Serio, S.

\bibitem[Molnar \& Birkinshaw (1999)]{MolnBirk99}
Molnar, S. M., \& Birkinshaw, M., 1999, \apj, 523, 78

\bibitem[Nagai et al. (2000)]{Nagaet00}
Nagai, D., Sulkanen, M. E., Evrard, A. E., 2000, \mnras, 316, 120

\bibitem[Peebles (1993)]{Peeb93}
Peebles, P. J., 1993, ``Principles of Physical Cosmology'', 
Princeton: Princeton University Press

\bibitem[Pen (1997)]{Pen_97}
Pen, U, 1997, New Astronomy, 2, 309

\bibitem[Perlmutter et al. (1999)]{Perlet99}
Perlmutter, S., et al., 1999, \apj, 517, 565

\bibitem[Peterson et al. (2001)]{Peteet01}
Peterson, J. R., et al., 2001, \aap, 365, L104

\bibitem[Petrosian (2001)]{Petr01}
Petrosian, V., 2001, astro-ph/0101145

\bibitem[Puy et al. (2000)]{Puy_et00}
Puy, D., Grenacher, L., Jetzer, Ph., and Signore, M., 2000, \aap, 363, 415

\bibitem[Reese et al. (2000)]{Rees00}
Reese, E. D., et al., \apj, 533, 38

\bibitem[Rephaeli (1995)]{Reph95} 
   Rephaeli, Y., 1995, \araa, 33, 541

\bibitem[Rephaeli \& Yankovitch (1997)]{RephYank97}
Rephaeli, Y., \& Yankovitch, D., 1997,\apjl, 481, L55

\bibitem[Riess et al. (2000)]{Rieset00}
Riess, A. G., et al. 2000, astro-ph/0001384

\bibitem[Rizza et al. (2000)]{Rizzet00}
Rizza, E., Loken, C., Bliton, M., Roettiger, K., \& Burns, J., 2000, \aj, 119, 21

\bibitem[Roettiger et al. (1995)]{Roetet95}
Roettiger, K., Burns, J. O., \& Pinkney, J., 1995, \apj, 453, 634

\bibitem[Roettiger et al. (1997)]{Roetet97}
Roettiger, K., Stone, J. M., \& Mushotzky, R. F., 1997, \apj, 482, 588

\bibitem[Sarazin (1999)]{Sara99}
Sarazin, C. L., 1999, \apj, 520, 529

\bibitem[Sarazin (1988)]{Sara88}
Sarazin, C. L., 1988, ``X-ray emissions from clusters of galaxies'', 
  Cambridge University Press, Cambridge

\bibitem[Sasaki (1996)]{Sasa96}
Sasaki, S., 1996, PASJ, 48, L119

\bibitem[Schlickeiser (1991)]{Schl91}
Schlickeiser, R., 1991, \aap, 248, L23

\bibitem[Schmidt et al. (2001)]{Schet01}
Schmidt, R. W., Allen, S. W., \& Fabian, A. C., 2001, astro-ph/0107311

\bibitem[Silk \& White (1978)]{SilkWhit78}
Silk, J, \& White, S. D. M., 1978, \apjl, 226, 103

\bibitem[Seljak (1996)]{Sel96}
Seljak, U., 1996, 463, 1

\bibitem[Sulkanen (1999)]{Sulk99}
Sulkanen, M. E., 1999, \apj, 522, 59

\bibitem[Sunyaev \& Zel'dovich (1980)]{sz80}  
Sunyaev, R. A., \& Zel'dovich, Y. B., 1980, \araa, 18, 537

\bibitem[Tamura et al. (2001)]{Tamet01}
Tamura, T., et al., 2001, \aa, 365, L87

\bibitem[Tegmark \& Zaldarriaga (2000)]{TegmZald00}
Tegmark, M. \& Zaldarriaga, M., 2000, astro-ph/0002091

\bibitem[Tegmark et al. (1998)]{Tegmet98}
Tegmark, M., Eisenstein, D. J., Hu, W., \& Knor, R. G., 1998, astro-ph/9805117

\bibitem[Tozzi \& Norman (2000)]{TozzNorm00}
Tozzi, P., \& Norman, C., 2000, astro-ph/0003289

\bibitem[Turner (2000)]{Turn00}
Turner, M. S., 2000, Physica Scripta, 85, 210

\bibitem[Ueda et al. (1993)]{Uedet93}
Ueda, H., Itoh, M., \& Suto, Y., 1993, \apj, 408, 3

\bibitem[White (1998)]{Whit98}
White, M., 1998, \apj, 506, 495

\bibitem[White \& Sarazin (1987)]{WhitSara87}
White, R. E., \& Sarazin, C. L., 1987, \apj, 318, 629

\bibitem[Yoshikawa et al. (1998)]{Yoshet98}
Yoshikawa, K., Itoh, M., \& Suto, Y., 1998, \pasj, 50, 203

\bibitem[Zaroubi et al. (1998)]{Zaroet98}
Zaroubi, S., Squires, G., Hoffman, Y., \& Silk, J., 1998, \apjl, 500, L87

\bibitem[Zaldarriaga et al. (1997)]{Zalet97}
Zaldarriaga, M., Spergel, D. N., \& Seljak, U., 1997, \apj, 488, 1


\end{thebibliography}
\end{document}